%% file: alma7465.tex
\documentclass[twocolumn]{aastex62}
\usepackage{verbatim}

\newcommand\hi{H{\small I}}
\newcommand\htoo{H$_2$}

\newcommand\cii{[C {\small II}]}

\newcommand\oiiihb{[\ion{O}{3}]/H$\beta$}
\newcommand\oiii{[\ion{O}{3}]}
\newcommand\niiha{[\ion{N}{2}]/H$\alpha$}

\newcommand\thirco{$^{13}$CO}
\newcommand\tweco{$^{12}$CO}
\newcommand\thirc{$^{13}$C}
\newcommand\twec{$^{12}$C}
\newcommand\ceighto{C$^{18}$O}
\newcommand\eighto{$^{18}$O}
\newcommand\sixo{$^{16}$O}
\newcommand\hcop{HCO$^+$}
\newcommand\chhhoh{CH$_3$OH}
\newcommand\cch{C$_2$H}
\newcommand\atlas{ATLAS$^{\rm 3D}$}
\newcommand\kms{km s$^{-1}$}
\def\arcsec{$^{\prime\prime}$}
\newcommand\solmass{M$_\odot$}
\newcommand\mstar{M$_\star$}
\newcommand\peryr{yr$^{-1}$}

\newcommand\mjb{mJy~bm$^{-1}$}
\newcommand\microjyb{$\mu$Jy~bm$^{-1}$}
\newcommand\jybkms{Jy~bm$^{-1}$ km~s$^{-1}$}
\newcommand\jykms{Jy~km~s$^{-1}$}
\newcommand\percc{~cm$^{-3}$}

\newcommand\e[1]{$\times 10^{#1}$}

\received{26 November 2020}
\revised{25 January 2021}
\accepted{27 January 2021}

\shorttitle{Molecular Gas in NGC~7465}
\shortauthors{Young et al.}

\begin{document}

\title{The Evolution of NGC~7465 as Revealed by its Molecular Gas Properties}

\correspondingauthor{Lisa Young}
\email{lisa.young@nmt.edu}

\author[0000-0002-5669-5038]{Lisa M.\ Young}
\affil{Physics Department, New Mexico Institute of Mining and Technology, 801 Leroy Place, Socorro, NM 87801, USA}
\affil{Adjunct Astronomer, National Radio Astronomy Observatory, Socorro, NM 87801, USA}
\author[0000-0001-9436-9471]{David S.\ Meier}
\affil{Physics Department, New Mexico Institute of Mining and Technology, 801 Leroy Place, Socorro, NM 87801, USA}
\affil{Adjunct Astronomer, National Radio Astronomy Observatory, Socorro, NM 87801, USA}
\author[0000-0003-4980-1012]{Martin Bureau}
\affil{Sub-department of Astrophysics, Department of Physics, University of Oxford, Denys Wilkinson Building, Keble Road, Oxford, OX1 3RH, UK}
\author[0000-0001-8513-4945]{Alison Crocker}
\affil{Department of Physics, Reed College, Portland, OR 97202, USA}
\author[0000-0003-4932-9379]{Timothy A.\ Davis}
\affil{School of Physics \& Astronomy, Cardiff University, Queens Buildings, The Parade, Cardiff, CF24 3AA, UK}
\author[0000-0003-2132-5632]{Sel{\c c}uk Topal}
\affil{Department of Physics, Van Y\"uz\"unc\"u Y{\i}l University, Van 65080, Turkey}

\begin{abstract}

We present ALMA observations of CO isotopologues and high-density molecular tracers (HCN, \hcop, CN, etc.) in NGC~7465, 
an unusually gas-rich early-type galaxy that acquired its cold gas recently.
In the inner 300 pc, the molecular gas kinematics are misaligned with respect to all other galaxy components; as the gas works its way inward it is torqued into polar orbits about the stellar kinematically-decoupled core (KDC),
indicating that the stellar KDC is not related to the current gas accretion event.
The galaxy also exhibits unusually high \tweco/\thirco\ line ratios in its nucleus but typical \thirco/\ceighto\ ratios. 
Our calculations show that this result does not necessarily indicate an unusual [\tweco/\thirco] abundance ratio but rather that 
\tweco\ (1-0) is optically thin due to high temperatures and/or large linewidths associated with the inner decoupled, misaligned  molecular structure.
Line ratios of the higher-density tracers suggest that the densest phase of molecular gas in NGC~7465 has a lower density than is typical for nearby galaxies, possibly as a result of the recent gas accretion.
All of the observed molecular properties of NGC~7465 are consistent with it having acquired its molecular (and atomic) gas from a spiral galaxy.  Further detailed studies of the CO isotopologues in other early-type galaxies would be valuable for investigating the histories of those that may have acquired their gas from dwarfs. 
Finally, these ALMA data also show an unidentified line source that is probably a background galaxy similar to those found at $z=1-3$ in blind CO surveys.

\end{abstract}


\section{Introduction} \label{sec:intro}

The gas in early-type galaxies (ETGs) offers interesting clues into their evolution, as there can be a much stronger disconnect between the stars and the gas than there is in spiral galaxies.  In spirals, the continuous star formation activity implies that there has been a continuous supply of cold gas over a Hubble time, and the stars and the gas have co-evolved in an uninterrupted symbiotic relationship.  In contrast, in early-type galaxies much of the cold and warm gas is kinematically misaligned (even retrograde) with respect to the stars, such that there is little possibility the stars formed from that gas or the gas was expelled from those stars.  

We expect irregular gas kinematics in the outer parts of galaxies where the cold gas (especially \hi) is more vulnerable to interactions and will retain signatures of disruption for many Gyr.  But such misalignments are also present in the inner few kpc of early-type galaxies, where we find that 20\% of the regular, relaxed CO disks have polar or retrograde rotation with respect to the stars in the host galaxy.  Thus, as much as 30\% to 50\%
of the molecular gas in early-type galaxies has been acquired recently from some external source such as an accreted dwarf galaxy \citep{davis_misalign}.  Similar results are found for ionized gas \citep{jin2016,bryant2019}.  Of the remainder, some of the relaxed, prograde molecular gas may have been resident in its host early-type galaxy for a long time, through the galaxy's transition to the red sequence \citep[e.g.][]{davis+bureau}.  Some of the molecular gas probably also originated in its current host, but made an extended detour through a hot phase in the galactic halo before condensing back into a cold phase \citep[e.g.][]{russell19,davis_massiv}.

These varied histories also have implications for the physical properties of the gas.
If the cold gas in an early-type galaxy was accreted in a minor merger or from a relatively pristine cold flow, the metallicity and isotopic abundance patterns in the molecular gas could reflect a significantly different star formation history and stellar initial mass function (IMF) from those that produced the current stellar populations in the galaxy \citep{davis+young}.  On the other hand, if the gas has been resident in the galaxy for a long time, gradually mixing with the mass loss material from evolved stars in the galaxy, we would expect a much tighter correspondence between the metallicity and isotopic abundances of the stars and the gas. 

In this context we have been undertaking detailed studies of the chemical and physical properties of the molecular gas in early-type galaxies; these chemical and physical properties might also serve as indicators of the galaxies' histories in a complementary manner to the information revealed by kinematics.
Here we present Atacama Large Millimeter/submillimeter Array (ALMA) observations of the early-type galaxy NGC~7465, which is known to have recently accreted its cold gas (Section \ref{why7465}).  We present \tweco\ (1-0) data at 0.\arcsec8 (110 pc) resolution and \tweco, \thirco, \ceighto, and several other higher-density molecular tracers
at 2\arcsec\ (280 pc) resolution (Section \ref{obs}).
We describe the millimeter continuum emission, compare it to the optical nebular emission line ratios, and estimate the ionized gas metallicity (Section \ref{cont}).
We also make comparisons of CO to stellar and ionized gas kinematics at matched resolution (Section \ref{gasdist}). 
Quantitative analysis of the molecular line ratios, their spatial variations, and comparisons to other galaxy types 
(Sections \ref{7465radial} - \ref{sec:disc}) reveal clues to the evolution of NGC~7465 and the broader context of galaxy evolution.

\section{About NGC~7465}\label{why7465} 

NGC~7465 was originally selected because it is a member of the \atlas\ sample of early-type galaxies \citep{cappellari_a3d1}, so there is a wealth of information about its stellar and gas content, including
two-dimensional maps of its stellar kinematics, 
ionized gas distribution and kinematics, stellar populations, star formation history, atomic
gas, and molecular gas \citep[][and references therein]{davor,a3dco,serra12,mcdermid2015}.
It is one of the more CO-bright members of the \atlas\ sample, so 
there is extensive information on its $^{12}$CO J=1-0 emission.  Its stellar mass is $\log (M_\star/M_\odot) = 10.4$ and its
adopted distance is 29.3 Mpc \citep{cappellari_a3d1}.

Though it is an early-type galaxy, NGC~7465 shows moderate levels of current star formation.
\citet{davis_sfr} estimated its total star formation rate from its 22\micron\ WISE flux, giving 1.37 $\pm\ 0.5$ \solmass~\peryr\ and a specific star formation rate SFR/\mstar\ $\approx (5.5 \pm 2) \times 10^{-11}$~\peryr.
These rates are low when compared to nearby spirals, placing NGC~7465 below the ``star formation main sequence" or SFR--\mstar\ relation \citep[e.g.][]{cluver2014} 
but high when compared to other early-type galaxies \citep{davis_sfr}.  Its depletion time (SFR/M$_{gas}$) is relatively long, at 7 Gyr.
Its radio and far-IR continuum emission imply a radio/FIR $q$ ratio that is also consistent with star formation activity \citep{nyland_sfr}.

\begin{figure}
\includegraphics[width=\columnwidth]{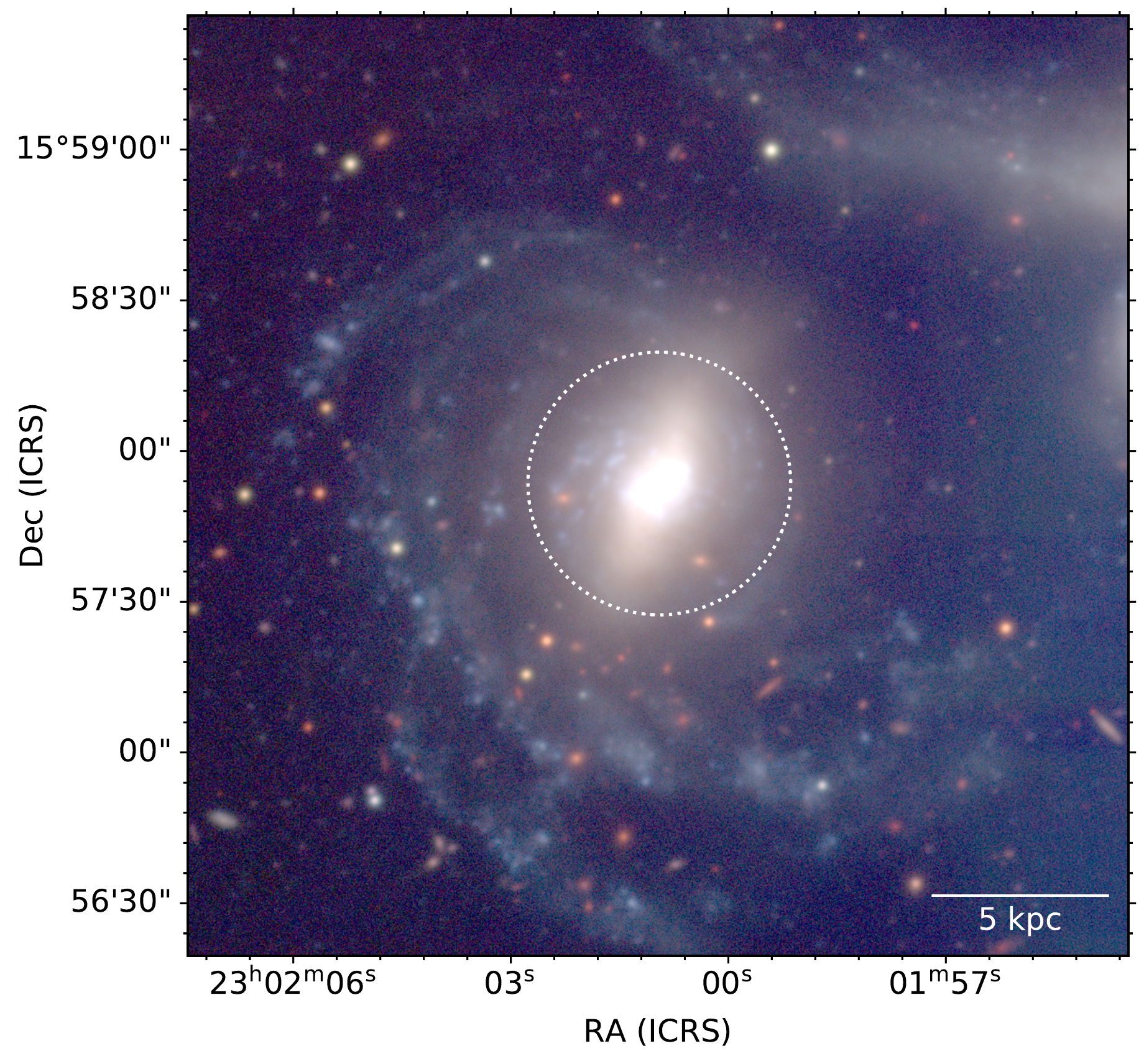}
\includegraphics[width=\columnwidth, trim=1cm 0.2cm 3cm 1cm, clip]{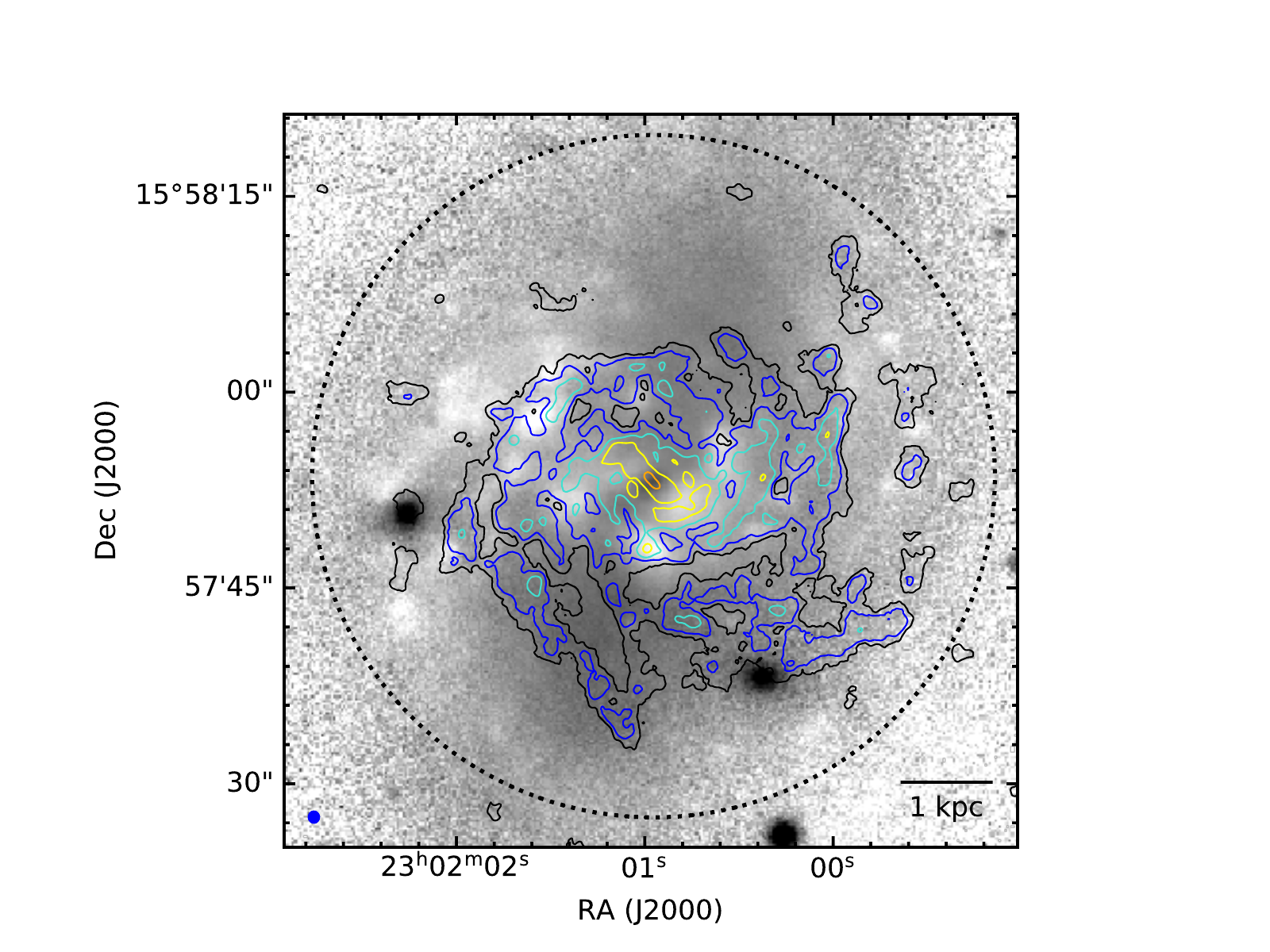}
\caption{Top: color composite of the MATLAS $u, g,$ and $i$ images \citep{duc15}.  Two neighboring galaxies are visible off the northwest corner of the image.  Bottom:  \tweco\ contours superposed on a $u-i$ color image (bluer colors are white).  Contour levels are 0.0164 $\times$ (1, 3, 9, 27, 81) \jybkms, where the lowest contour is the nominal column density sensitivity equivalent to a 5$\sigma$ signal in one channel of width 15 \kms.  The synthesized beam (0.\arcsec8 $\times$ 0.\arcsec7) is shown as a blue ellipse in the bottom left corner.  A primary beam correction has been applied; in both panels the dotted circle indicates the half power point of the ALMA primary beam at 114 GHz.}
\label{fig:7465mom0}
\end{figure}

In addition to being CO-bright, NGC~7465 is notable for signatures of interactions with near neighbors NGC~7463 and NGC~7464 at projected separations $<$ 20 kpc.  Though it is an early-type galaxy (an S0 galaxy and a fast rotator), it has faint blue outer arms at radii of 8 to 10 kpc (Figure \ref{fig:7465mom0}).  Its \hi\ emission is highly disturbed and strongly misaligned with respect to the stellar body of the galaxy \citep{li+seaquist,serra12}.  This kinematic evidence strongly suggests that the galaxy's cold gas was recently acquired from some external source, and this accretion/interaction event probably drove the formation of the blue outer arms.  The stellar kinematics show a kinematically-distinct core \citep{davor} which, as we will show, is probably not related to the most recent gas-transfer interaction.  A deep $u$-band image from \citet{duc15} also shows recent star formation activity in an irregular spiral structure at radii $\approx$ 3\arcsec\ to 20\arcsec\ ($\approx$ 0.4 to 3 kpc). 

NGC~7465 is classified as a barred S0 galaxy by some \citep[e.g.][]{rc3}, and indeed Figure \ref{fig:7465mom0} might be interpreted as a nearly face-on galaxy with a bar $\approx$ 1\arcmin\ (8.5 kpc) in length.  An alternative interpretation arising from the \atlas\ dynamical analysis \citep{cap:a3dJAM} is that the inner $r \lesssim 30$\arcsec\ of the galaxy is an edge-on axisymmetric fast rotator and the outer blue arms are transient extraplanar structures.  The primary reason for this alternative interpretation is that
the inner part of NGC~7465 does not show typical kinematic features of strongly barred early-type galaxies, which are (1) large misalignments between the photometric major axis of the bar and the kinematic major axis of the stars in that region, and (2) cylindrical rotation in the stellar kinematics, yielding parallel iso-velocity contours \citep{davor,a3d_bars}. 
 Additional stellar kinematic data covering a larger field of view would be helpful in distinguishing between these alternatives.

Finally, IRAM 30m observations of molecular lines in the \atlas\ galaxies  
have shown that NGC~7465 has some uncommon integrated molecular line ratios \citep{crocker_hd}.
Relative to \tweco, NGC~7465 has unusually faint \thirco\ and HCN, and it has a low HCN/\hcop\ ratio.
At higher resolution, these line ratios might reflect the influence of an active galactic nucleus on its surrounding medium.  The galaxy contains a Seyfert or LINER nucleus \citep[e.g.][]{goncalves99}, a 5 GHz synchrotron point source \citep{nyland_nuclei} and a {\it Swift} hard X-ray source \citep{baumgartner13}.
Our new ALMA data, with additional lines and higher spatial resolution, provide a better perspective on how to interpret the molecular line ratios and connect them to the physical properties of the gas and to galaxy evolution.

\section{Observations}\label{obs} 

 NGC~7465 has been observed by ALMA in several projects.  Here we used the Band 3 observations of \thirco, \ceighto, and CS in project 2018.1.01253.S (December 2018); \tweco, HCN, \cch, and \hcop\ in 2016.1.01119.S (November 2016); and \tweco\ and CN in 2018.1.01599.S (October 2018).
We used the standard pipeline-calibrated raw data and carried out continuum subtraction, imaging, and cleaning ourselves.  An additional round of phase self calibration was not found to offer any improvement in the images because of the relatively modest signal-to-noise ratios.
For continuum subtraction we used zero-order or first-order fits to the line-free channels in the visibility domain; for imaging, we made a wide variety of images at varying channel widths and resolutions as necessary to optimize the resolution or to match resolutions for resolved line ratios.  The times on source, along with fiducial (``natural" weighting) beam sizes and rms noise levels, are indicated in Table \ref{tab:rms2} for the detected spectral lines.  

Line emission in the data cubes was cleaned down to about the 1$\sigma$ level and primary beam corrections were applied.
The \tweco\ integrated intensity images (e.g.\ Figure \ref{fig:7465mom0}) were created by modestly smoothing the data cube in spatial and velocity dimensions and clipping on the smoothed cubes to create a velocity-dependent mask defining the volume with emission.  These \tweco\ masks were then applied to the fainter lines for creating their integrated intensity images.

Line fluxes reported in this paper include uncertainties which are only based on the thermal fluctuations in the data, representing signal-to-noise considerations; they do not include uncertainties in the absolute flux calibration scale, such as those due to secular variability in the calibrator sources.  The flux calibration of ALMA data that have been processed with standard calibration procedures is usually accurate to $\approx$ 5\% to 10\% \citep[e.g.][]{martin2019,andrews2018,almahandbook}. Line ratios between data taken in different observing setups and different tunings, such as \tweco/\thirco\ (see Table \ref{tab:rms2}), have an additional uncertainty due to this absolute flux calibration.  However, line ratios between data taken simultaneously using the same setup (e.g.\ \thirco/\ceighto\ or HCN/\hcop) do not have this additional uncertainty.

In addition to the spectral lines mentioned above, 
we also have upper limits on \chhhoh ($2_k-1_k$), SiO(2-1; $v=0$), and HNCO$(4_{0,4}-3_{0,3})$.
Furthermore, there is an unidentified line source in the vicinity of NGC~7465.  It seems to be a galaxy at a much higher redshift, and it is described in more detail in Appendix \ref{app:ufo}.

\begin{deluxetable}{lrlcc}
\tablecaption{Fiducial data cube parameters for NGC~7465\label{tab:rms2}}
\tablehead{
\colhead{Line} & \colhead{Time} &
\colhead{Beam size} &
\colhead{rms} & \colhead{$\Delta v$}\\
\colhead{} & \colhead{(sec)} & \colhead{(\arcsec)} &
\colhead{(\mjb)} & \colhead{(\kms)}
}
\startdata
\tweco(1-0) & 1590 & 1.60$\times$1.18 & 1.59 & 5.1 \\
                   & 11733 & 0.81$\times$0.74 &  0.53  & 2.6 \\
CN(1-0)      & 11733  &  0.84$\times$0.76 & 0.066  & 45.0 \\
\thirco(1-0) & 3810 & 1.89$\times$1.81 & 0.38 & 13.3\\ 
\ceighto(1-0) & 3810 & 1.90$\times$1.82 & 0.35 & 13.3\\  
CS(2-1) & 3810 & 2.12$\times$2.04 & 0.29 & 14.9 \\ 
\chhhoh ($2_k-1_k$) & 3810 & 2.18$\times$2.06 & 0.35 & 15.1 \\
\hcop(1-0) & 8890 & 1.70$\times$1.37 & 0.26 & 13.1 \\
HCN(1-0) & 8890 & 1.71$\times$1.38 & 0.26 & 13.2 \\
HNCO $(4_{0,4}-3_{0,3})$ & 8890 & 1.72$\times$1.39 & 0.28 & 13.3\\
\cch(N=1-0) & 8890 & 1.72$\times$1.38 & 0.29 & 13.4 \\
SiO(2-1, $v=0$)   & 8890 & 1.73$\times$1.39 & 0.28  & 13.5\\
\enddata
\tablecomments{Beam sizes and rms noise levels refer to images made with ``natural" uv-weighting and the listed channel widths.  In the case of the \tweco\ and CN observations, the channel widths in the last column are close to the best velocity resolutions allowed by the data.  For the other lines, the listed channel widths are representative but not the best possible.   At a distance of 29.3 Mpc, 1\arcsec\ corresponds to 140 pc.}
\end{deluxetable}

\section{Radio continuum and ionized gas}\label{cont} 

\begin{figure}
\includegraphics[width=\columnwidth, trim=1cm 0.3cm 1.5cm 1cm, clip]{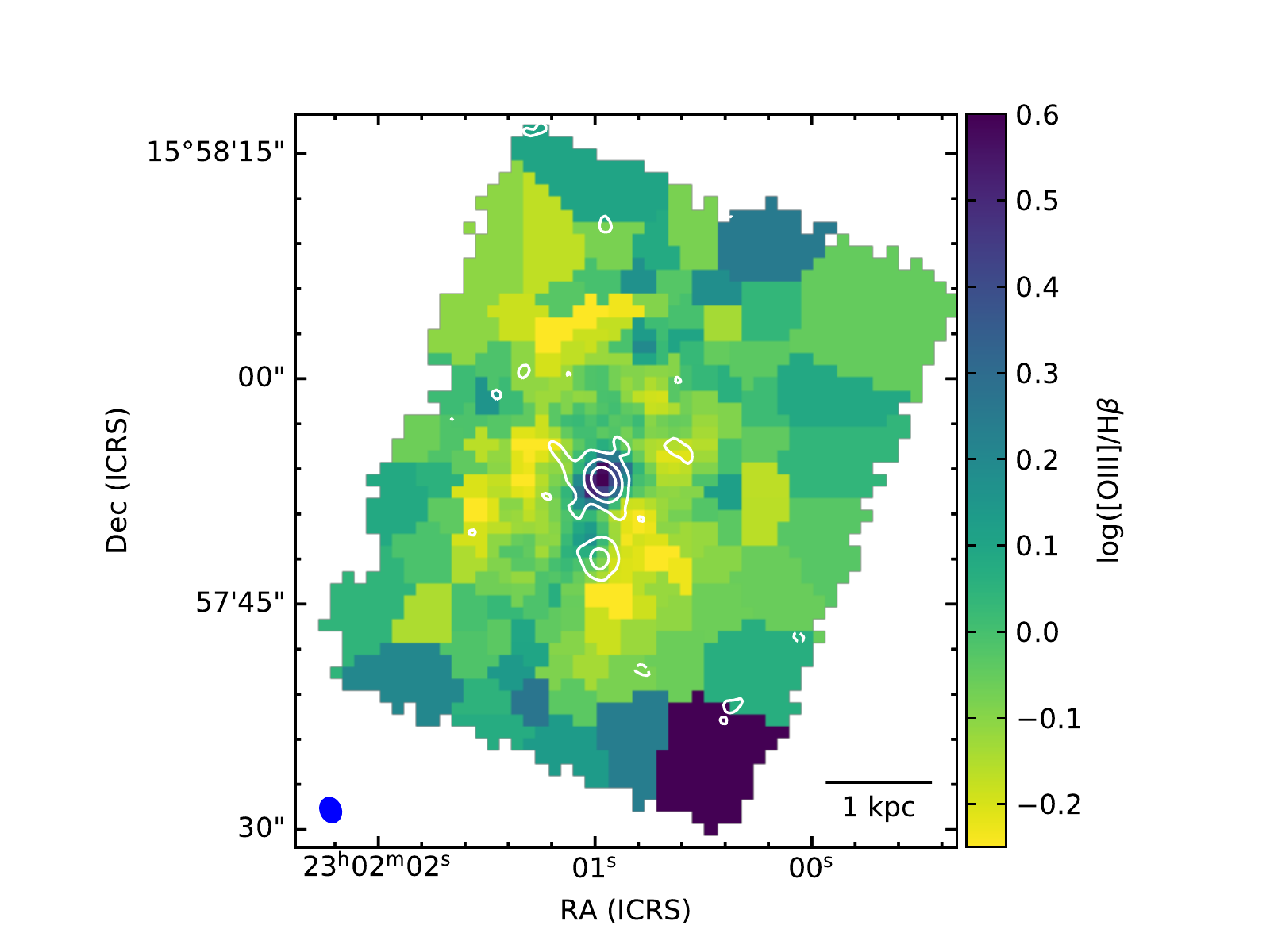}
\caption{The \oiiihb\ ratio from the \atlas\ project \citep{cappellari_a3d1}.  Contours are the 3mm continuum emission, showing the two point sources.  Contour levels are ($-3$, 3, 10, and 25) times the rms noise, which is 11 \microjyb; the peak intensity is 590 \microjyb.  The beam size of the continuum emission is shown as the solid blue ellipse in the lower left corner.
\label{fig:7465cont}}
\end{figure}

As we have observations at similar angular resolutions covering a large range of frequencies from 86 to 115 GHz (with gaps), 
the data are suitable for imaging the continuum intensity and estimating the spectral index of any emission.
NGC~7465  has two continuum point sources, one in the nucleus and one 5\arcsec\ south of the nucleus (Figure \ref{fig:7465cont}).  The nuclear source has a falling spectral index; its flux densities measured at 86.15 and 107.9 GHz are 742 $\pm$ 12 $\mu$Jy and 616 $\pm$ 18 $\mu$Jy, respectively.  Parametrizing the flux density as $S_\nu \propto \nu^\alpha$ therefore gives $\alpha = -0.83 \pm 0.13.$  Similarly, using
all of the available line-free frequencies in CASA's multi-term, multi-frequency synthesis deconvolver \citep{rau_mfs}, 
we find $\alpha = -0.59 \pm 0.12.$  These estimates are consistent with each other and both clearly indicate synchrotron emission in the nucleus of the galaxy, which is not surprising given the other active galactic nucleus (AGN) indicators mentioned in Section \ref{why7465}.  For comparison, \citet{noema7465} find that at 230 GHz the nuclear continuum source has a flux density of 900 $\pm$ 200 $\mu$Jy, which is broadly consistent with the flux densities measured here and suggests that in this mm regime the nuclear source's spectrum is flattening as dust emission begins to dominate.

The fainter continuum source has flux densities of 155 $\pm$ 12 $\mu$Jy and 245 $\pm$ 18 $\mu$Jy at 86.15 and 107.9 GHz, suggesting instead a rising spectral index consistent with dust emission. 
It appears to be associated with 
a  region of recent star formation activity, as indicated by a bright blue source in the $u$ image, peaks in \oiii\ and H$\alpha$ emission \citep{ferruit}, and 
a small peak in the molecular surface brightness.

\citet{moustakas06} presented optical spectroscopy and measurements of the nebular emission line fluxes in NGC~7465.  The nebular spectrum (from a region 2.\arcsec5 square) has log(\niiha) = -0.28 $\pm$ 0.03 and log(\oiiihb) = 0.44 $\pm$ 0.03, suggesting the ionization there is dominated by the AGN 
\citep{moustakas10}.
The spectrum integrated over a much larger 50\arcsec\ $\times$ 60\arcsec\ rectangular region has log(\niiha) = -0.39 $\pm$ 0.03 and log(\oiiihb) = 0.06 $\pm$ 0.03, consistent with star formation activity.
The \atlas\ data provide more spatial information at higher resolution, though only on the \oiiihb\ ratio; they show an even higher ratio of log(\oiiihb) = 0.62 in the nucleus, on the location of the nuclear radio synchrotron and hard X-ray sources (Figure \ref{fig:7465cont}).   Lower ratios of log(\oiiihb) from $-0.25$ to $-0.3$ are seen in patches at radii of 3\arcsec\ to 10\arcsec, and while this single ratio cannot conclusively identify the source of ionization \citep[e.g.][]{sarzi_ioniz}, all the data are consistent with the interpretation that the current star formation activity in the interior of the galaxy is particularly concentrated in this annular region.

The nebular emission line fluxes in \citet{moustakas06} also enable an estimate of the gas metallicity in NGC~7465.
We use line fluxes integrated over the large 50\arcsec\ $\times$ 60\arcsec\ region; 
following the prescriptions outlined in \citet{moustakas10}, the emission line ratios suggest a metallicity consistent with solar. 
We find 12 + log(O/H) = 8.38 $\pm$ 0.06 using the calibration of Pilyugin \& Thuan (2005) and $\approx$ 9.0 using the calibration of Kobulnicky \& Kewley (2004); each of these is consistent with its respective median of the SINGS galaxies (mostly spirals) in \citet{moustakas10}.

\section{Molecular Gas Distribution and Kinematics}\label{gasdist} 

\subsection{Large-scale disk}

The \tweco\ emission from NGC~7465 (Figure \ref{fig:7465mom0}) shows a disky structure consisting of irregular, flocculent spiral arms extending to radii of at least 20\arcsec.  Some emission is also detected beyond the half-power point of the ALMA primary beam, so it seems likely that there is significant molecular gas beyond the region that we are able to image in these datasets.
This irregular disk has an axis ratio near 1, suggesting a low inclination, but its kinematic major axis is strongly twisted and asymmetric as might be expected of recently accreted gas (Figure \ref{fig:7465mom1}).
Although there is a large mismatch between the angular resolution of the \hi\ and CO data, the CO velocity field on large scales is consistent with that of the \hi\ \citep{li+seaquist}, suggesting that those two gas phases probably have a common origin.

Section \ref{why7465} discusses the fact that it is not clear whether NGC~7465 is a face-on barred galaxy or an edge-on axisymmetric galaxy with extraplanar arm/ring structures.
In either case, steady-state models are
likely to be misleading when trying to interpret this galaxy's kinematics.  The interactions with its neighbors mean that its potential is rapidly time-varying.  Either these interactions, or the perturbations caused by a bar (if one is present), might explain the non-circular motions and possible gas inflows discussed in this section. 
In addition, we note that dense molecular gas is found over most of the extent of the elongated structure.  In a bar model, this would imply that either dense gas is present beyond the inner Lindblad resonance (ILR), contrary to expectation \citep[e.g.][]{athanassoula92}, or alternatively that the radius of the ILR is much larger than is usual in early-type disk galaxies and it extends to almost the end of the bar and thus corotation. 
Overall, we therefore argue that it is problematic to interpret the gas distribution and kinematics in NGC~7465 in terms of a relaxed/steady state early-type barred disk galaxy potential.

\subsection{Inner kpc and decoupled core}

\begin{figure*}
\includegraphics[width=\textwidth, trim=5mm 2cm 2cm 1cm, clip]{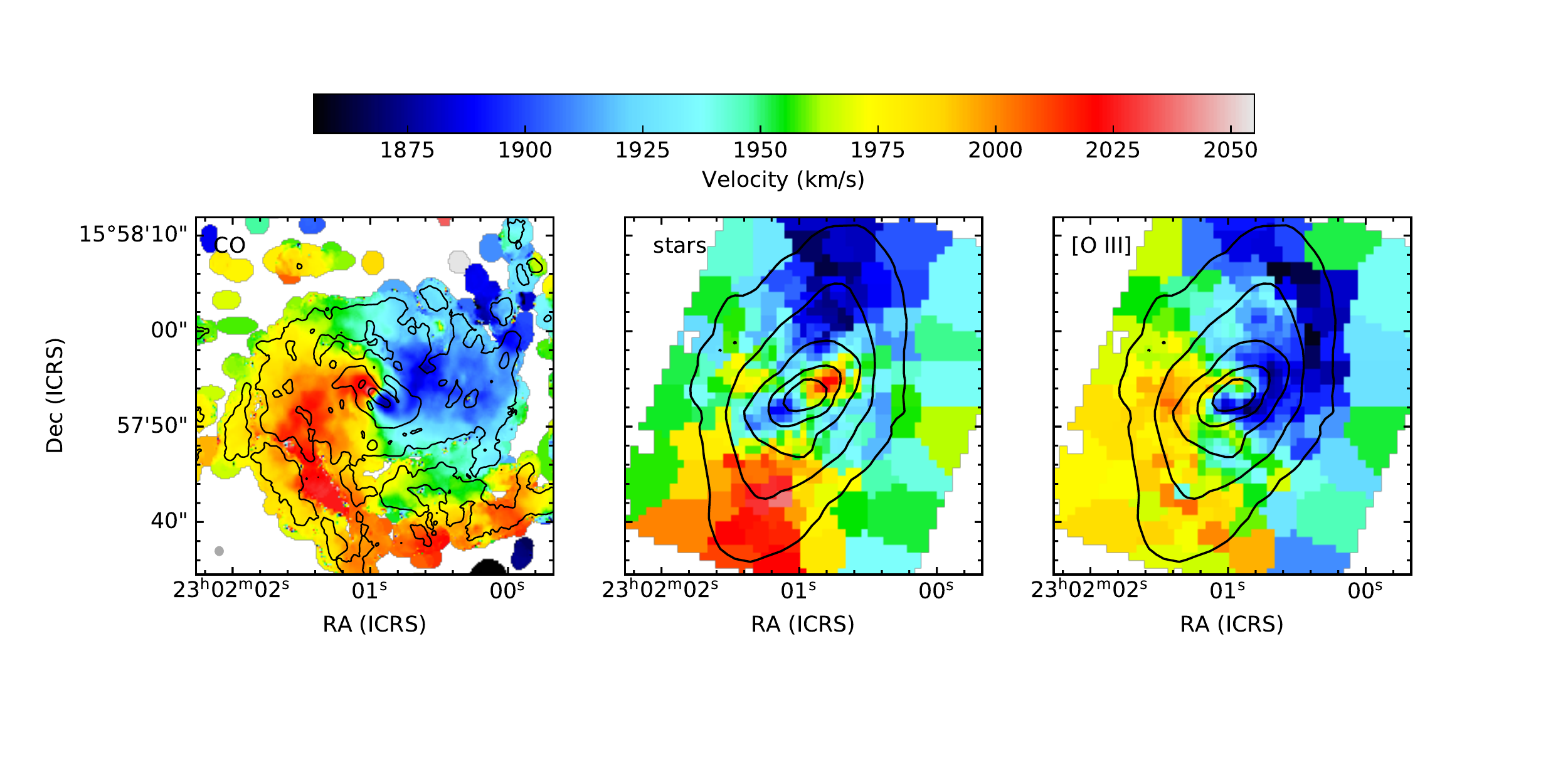}
\caption{CO, stellar, and ionized-gas velocity fields of NGC~7465. In the \tweco\ velocity field, the contours show the integrated CO intensity at 0.8\arcsec $\times$ 0.7\arcsec\ resolution, with contour levels at 0.03, 0.3, and 1.0 \jybkms.  
The CO beam size is shown as a gray ellipse in the bottom left corner.
These CO data have slightly better angular resolution than the ground-based optical data, which are from \citet{cappellari_a3d1}.  In the stellar and ionized gas panels, the contours are stellar isophotes.
\label{fig:7465mom1}}
\end{figure*}

At radii $\lesssim$ 5\arcsec\ (where we have most of our molecular line detections) the \atlas\ stellar isophotes are flattened and their major axis is aligned to the kinematically-decoupled stellar core (KDC; Figure \ref{fig:7465mom1}, center panel).  The KDC is both photometrically and kinematically misaligned by 150\arcdeg\ with respect to stellar structures farther out.  In this region the molecular gas forms a bright ridge about 10\arcsec\ long (1.4 kpc), oriented along a northeast-southwest axis.  The ridge is misaligned by about 100\arcdeg\ with respect to the stellar KDC and 120\arcdeg\ with respect to the stellar kinematic and photometric axes at larger radii.
The molecular gas shows a strong velocity gradient along the ridge, so that the most extreme CO velocities are found along the ridge, about 1\arcsec\ from the nucleus as defined by the radio continuum position.
\tweco\ (2-1) emission at higher resolution \citep{noema7465} shows a patchy spiral feature that can be followed inward to a radius of 2\arcsec\ and a smaller molecular ridge 2\arcsec\ long, with a saddle at the location of the AGN.  But the smaller ridge observed in (2-1) emission maintains the same orientation as the more extended ridge we observe in the present data; thus, on scales of 1\arcsec\ (140 pc) the molecular ridge is still perpendicular to the stellar KDC.

\citet{davis_misalign} discussed the kinematics of ionized and molecular gas in early-type galaxies, and noted that the two phases agreed in all of the cases they studied at moderate ($\approx$ 1 kpc) resolution.  That is also still true in NGC~7465 for radii $\gtrsim$ 2\arcsec\ (280 pc), but for smaller radii the kinematics of the ionized and molecular gas diverge.  The ionized gas continues its inward twist such that at small radii the kinematic position angle of the ionized gas nearly matches that of the stellar KDC but is 45\arcdeg\ offset from the molecular kinematic position angle.  The molecular gas, on the other hand, maintains its fixed orientation nearly perpendicular to the stellar KDC.  This divergence of the kinematics is suggestive of more complex noncircular motions or perhaps multiple ionization sources in the galaxy's nucleus, such that the inner ionized gas is decoupled from the molecular gas.  The ionized gas kinematics in the center of NGC~7465 may even be tracing an AGN-driven ionized outflow, and Appendix \ref{app:outflow} presents a closer look at that evidence.

The mismatch between molecular and stellar kinematics in the center of NGC~7465 also means that the stellar KDC has not formed out of the molecular gas currently present in the galaxy.  It must represent a previous formation event, or perhaps a much earlier episode of an extended interaction.
More detailed observations of the stellar populations in the KDC would be required to date its formation \citep[e.g.][]{sarzi16}.

\subsection{Other molecular species in the inner kpc}

\begin{figure*}
\includegraphics[width=\textwidth, trim=0.5cm 0.5cm 1cm 0cm, clip]{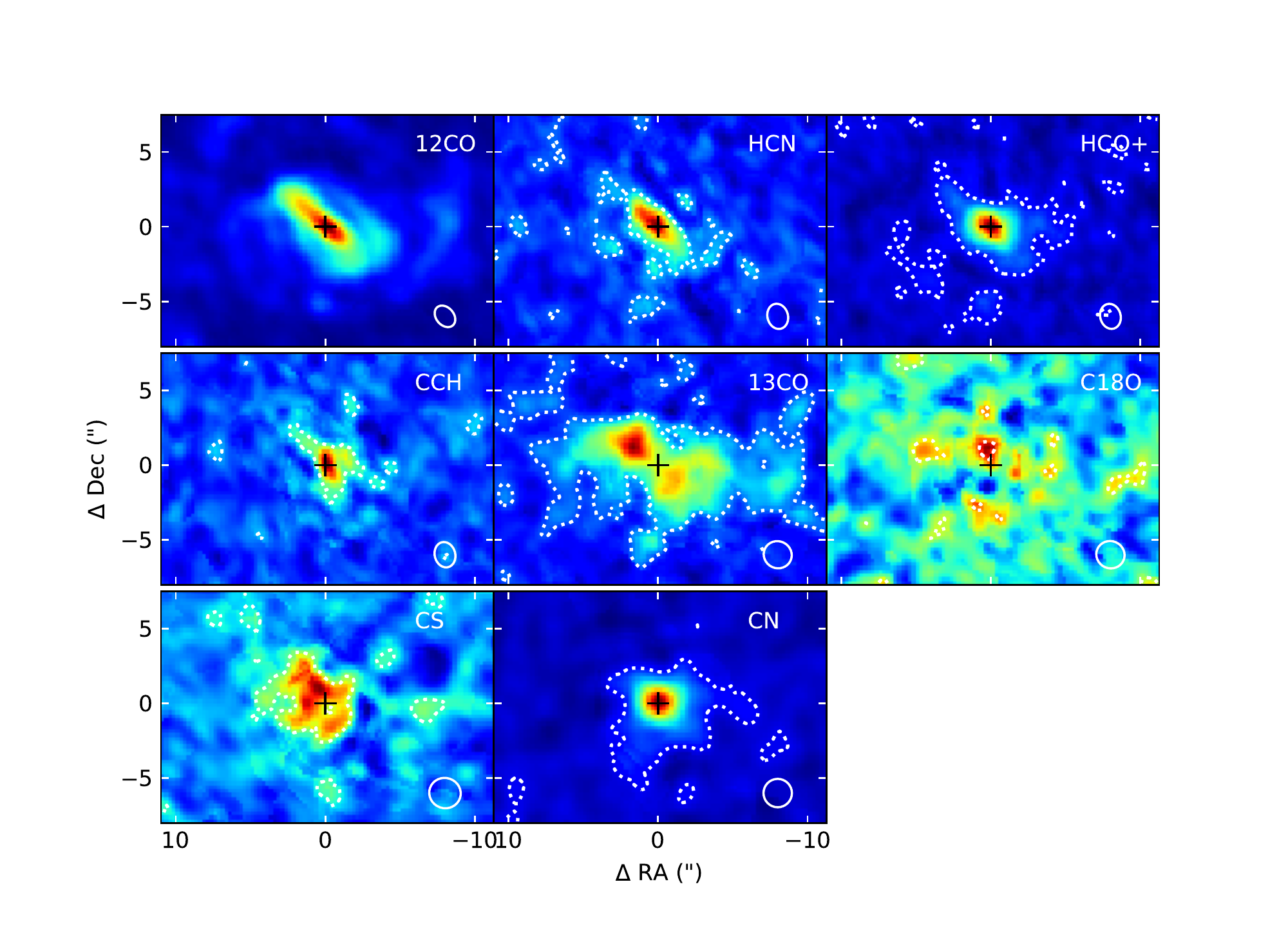}
\caption{Integrated intensity images of multiple molecular species in the inner few kpc of NGC~7465.  The position of the nuclear 3mm continuum source is marked with a cross, to guide the eye.  These integrated intensities are computed using a velocity-dependent mask that follows the rotation of the gas, as defined by  \tweco\ emission.  Each line's synthesized beam is indicated by the ellipse in the lower right corner, and the colors are scaled to the minimum and maximum of each image.  Integrated line fluxes towards the nucleus are given in Table \ref{tab:coldens}.  For the fainter lines, a dotted white contour indicates pixels where the integrated intensity has a signal-to-noise ratio $> 2.5.$  The corresponding \tweco\ contour is mostly outside this field of view.\label{fig:7465m0s}}
\end{figure*}

The integrated intensity maps of the other molecular lines are presented in Figure \ref{fig:7465m0s}.  To avoid biases related to the different signal-to-noise ratios, a velocity-dependent mask tracing the \tweco\ emission is used for all of the other spectral lines.  Their distributions can be broadly classified into two major groups.  \tweco, HCN, \hcop, CN, and \cch\ are centrally concentrated and peaked on the nucleus of the galaxy, whereas \thirco\ and \ceighto\ show peaks a couple of arcseconds ($\approx$ 1 beam) north and east of the nucleus rather than on the nucleus itself.
In \thirco, for example, the integrated intensity at the position of the nucleus is a factor of two lower than farther out along the molecular ridge.  CS emission is relatively weak and while it appears to have a peak off the nucleus, the spatially-resolved ratios (Section \ref{7465radial}) suggest that this appearance may be due to noise in the integrated intensity.  Finally, HNCO, \chhhoh\ and SiO emission are not detected in these data.

\begin{figure}
\includegraphics[width=\columnwidth, trim=0.4cm 0.3cm 1.5cm 1cm, clip]{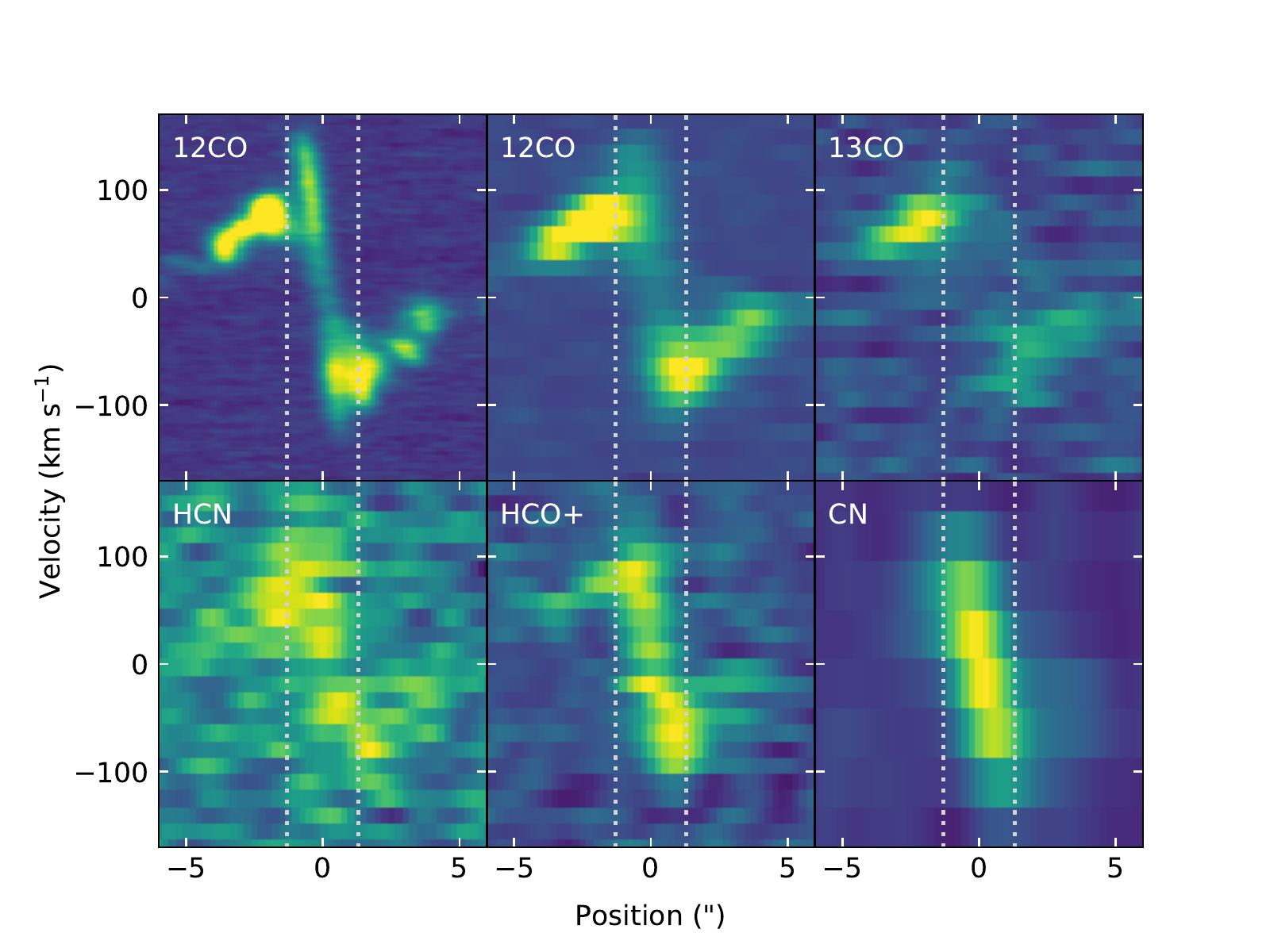}
\caption{Position-velocity slices at a position angle of +42\arcdeg, one beam wide.  The slices cut through the nucleus and follow the molecular ridge, tracking the kinematic major axis at small radii (see Figure \ref{fig:7465mom1}).  Velocities are measured with respect to the systemic velocity, 1961 \kms.  Dotted lines at $\pm$ 1.\arcsec3 demarcate the inner kinematic component with the strong velocity gradient. \tweco\ is plotted both at the highest available resolution and at a resolution matching the \thirco\ data.\label{fig:7465pvs}}
\end{figure}

Position-velocity slices through the data cubes (Figure \ref{fig:7465pvs}) are consistent with those from \citet{noema7465}, showing the presence of two kinematic structures in the central molecular gas of NGC~7465.  
At radii $<$ 1.\arcsec3 we find a nuclear ring or bar (or possibly an outflow) that is most prominent in \hcop\ and CN emission, also visible in HCN and \tweco\ emission, and extremely weak or absent in \thirco\ emission.  Exterior to 1.\arcsec3 the primary structure is the warped disk, where the \thirco\ emission is at its brightest, and there the \thirco\ emission follows the \tweco\ emission.  The appearance of a Keplerian decline in the \tweco\ emission in Figure \ref{fig:7465pvs} is an artifact of the twist in the kinematic position angle, so that at large radii this slice is closer to the kinematic minor axis than the major axis.

\section{Line strengths: CO isotopologues}\label{7465radial} 

\begin{figure}
\includegraphics[width=\columnwidth, trim=0.5cm 0.4cm 1cm 0.5cm, clip]{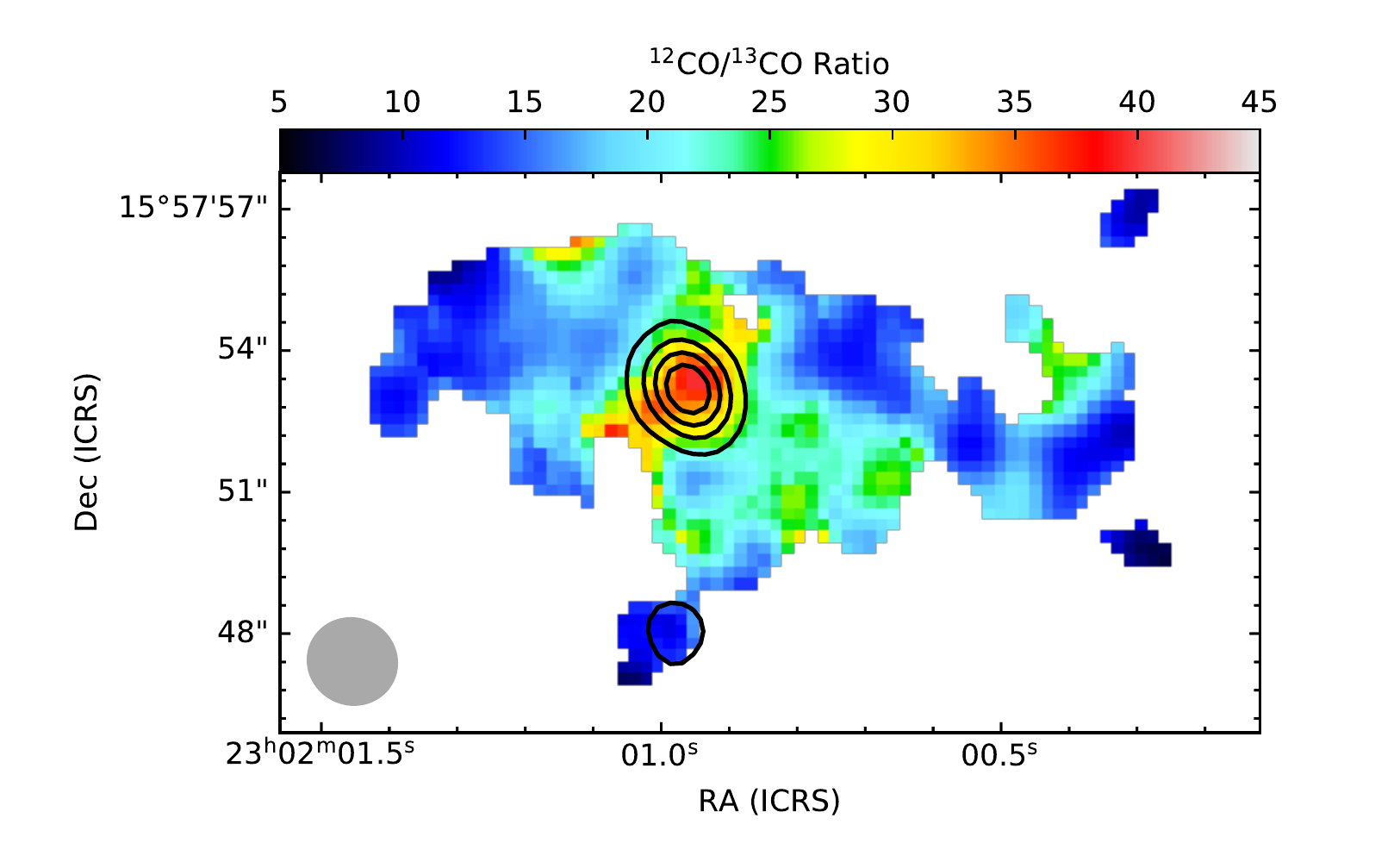}
\caption{\tweco/\thirco\ integrated intensity ratio (computed in units of brightness temperature).  Black contours show the 3mm continuum emission, and the angular resolution of the \tweco/\thirco\ ratio image is shown by the gray ellipse in the lower left corner.\label{fig:7465_1213}}
\end{figure}

\begin{figure}
\includegraphics[width=\columnwidth, trim=1.5cm 12.8cm 3cm 6cm,clip]{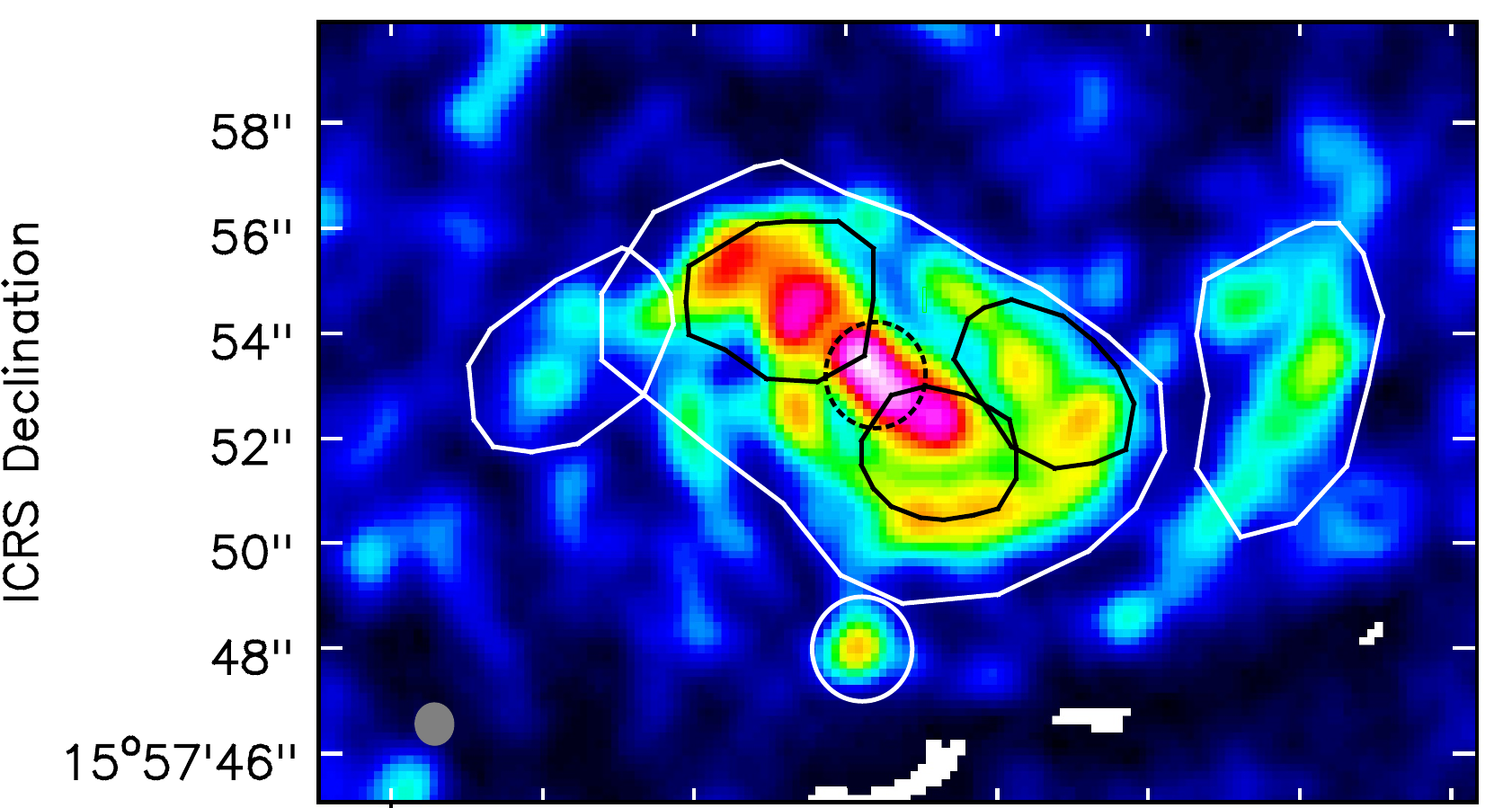}
\includegraphics[width=\columnwidth, trim=1.5cm 11cm 3cm 6.3cm,clip]{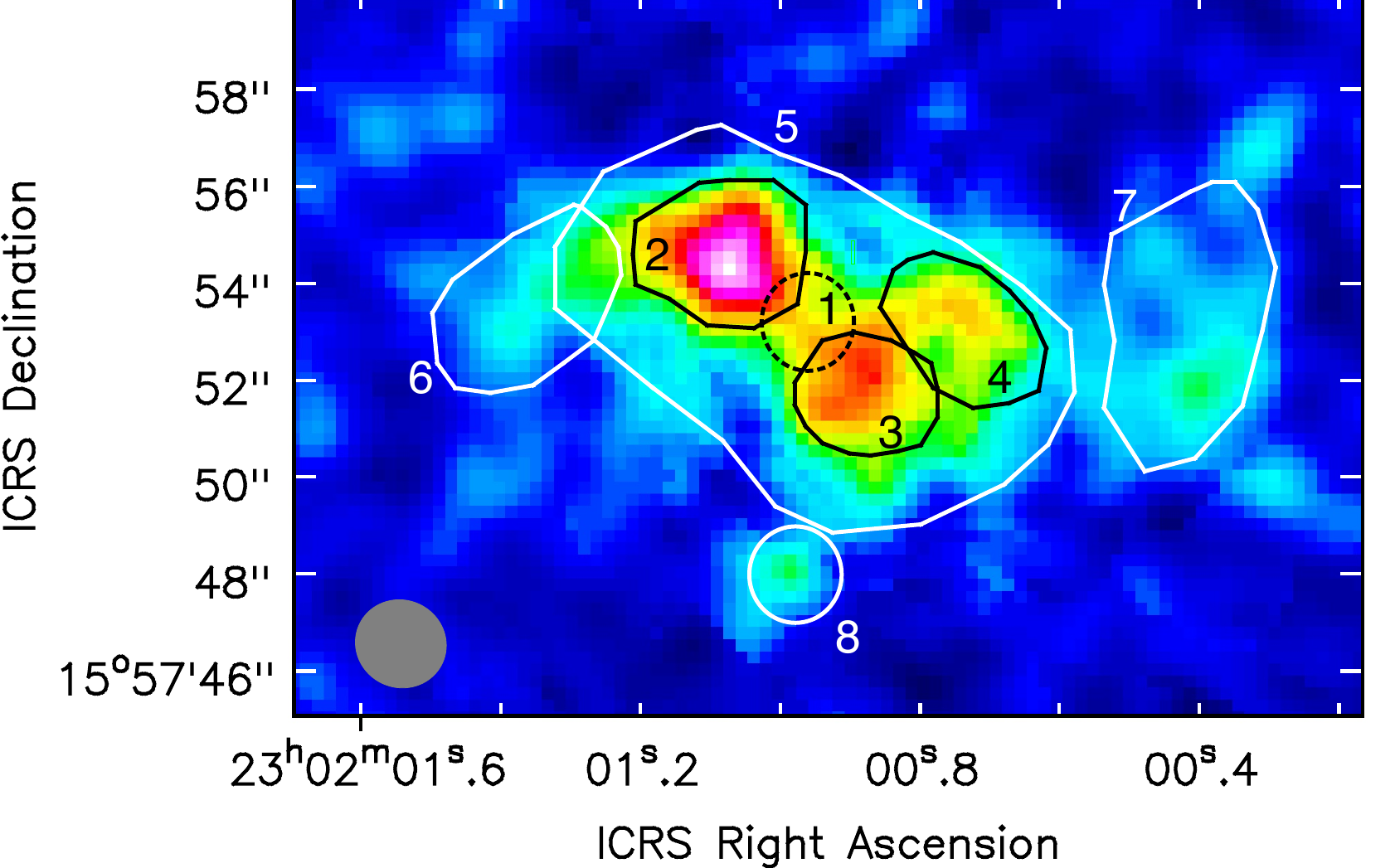}
\caption{Integrated intensities of \tweco\ (top) and \thirco\ (bottom).  The angular resolution in each panel is shown with a gray ellipse in the lower left corner.  Black and white polygons show the regions used for summing spectra, to investigate the spatial variations of the line ratios.  The boundary of the nuclear region is dotted.  A few white pixels in the lower portion of the top panel are locations where \tweco\ is not detected at this angular resolution.  Numbers in the lower panel identify the regions for reference to the table in Appendix \ref{app:bigtab}.
\label{fig:regions}}
\end{figure}

\begin{figure*}
\centering
\includegraphics[width=\textwidth, trim=0.5cm 0cm 0cm 0cm]{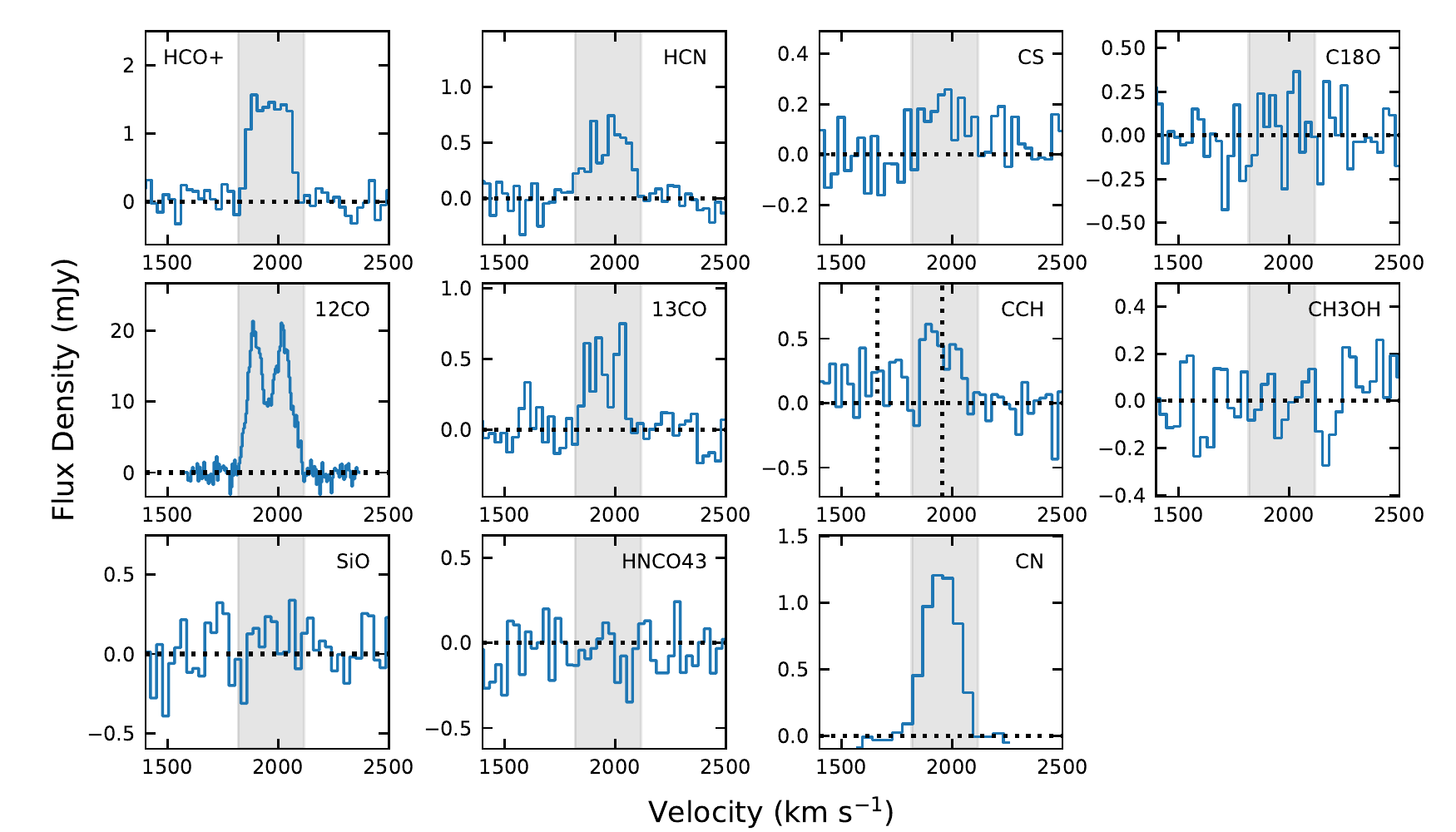}
\caption{Spectra extracted from the nuclear region, which is 2\arcsec\ (1 beam) in diameter and centered on the 3mm continuum point source.  The grey boxes mark the velocity range used for integrating at this position.  In the \cch\ panel, the two dotted lines mark the centers of the two main fine structure blends and the velocity scale is calculated for the low frequency ($J= \frac{3}{2} - \frac{1}{2}$) blend.  These nuclear spectra exhibit unusually weak \thirco\ emission.
\label{fig:7465lines}}
\end{figure*}

\begin{figure}
\includegraphics[width=\columnwidth, trim=4mm 4mm 13mm 0cm, clip]{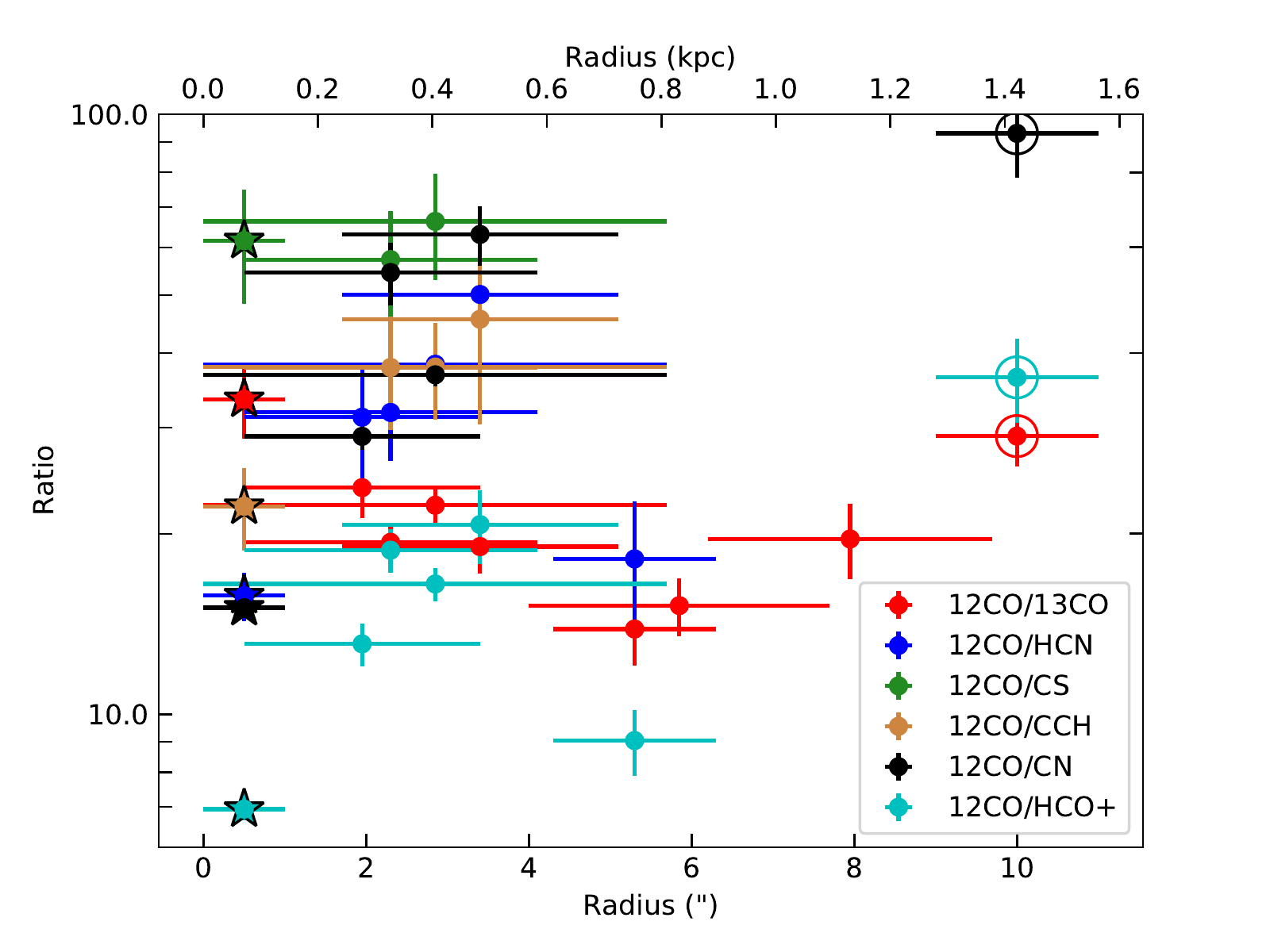}
\caption{Line ratios are integrated over the regions shown in Figure \ref{fig:regions}; the horizontal error bars indicate the range of radii probed by each region.  Ratios measured towards the nucleus are plotted as stars, and those measured over the entire region with detected \tweco\ emission are plotted at 10\arcsec\ and circled.  The region with relatively low \tweco/HCN at 5.\arcsec3 is centered on the fainter 3mm continuum source.  For purposes of clarity, only the detections ($>3\sigma$) are plotted.\label{7465radial1}}
\end{figure}

\begin{figure}
\includegraphics[width=\columnwidth, trim=4mm 4mm 13mm 0cm, clip]{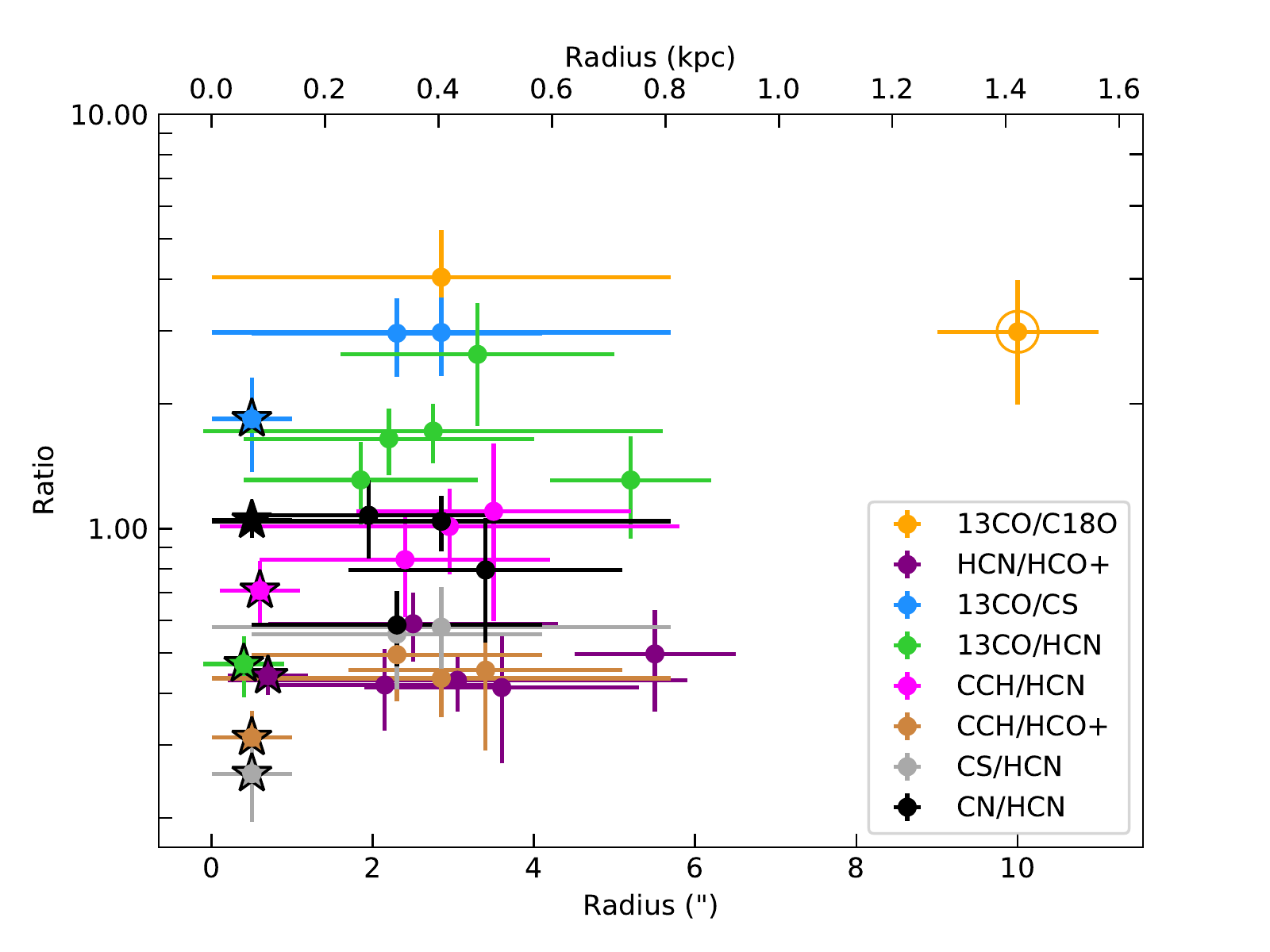}
\caption{Similar to Figure \ref{7465radial1},  for different ratios; symbols have the same meanings.  The radial coordinates for some ratios have been slightly shifted for better visibility.\label{7465radial2}}
\end{figure}

For the bright lines of NGC~7465, we can create line ratio maps from the integrated intensity images, as in Figure \ref{fig:7465_1213} for \tweco/\thirco.  For the fainter lines we measure integrated line fluxes and ratios by defining spatial regions (Figure \ref{fig:regions}), summing the cubes within the spatial regions to produce integrated spectra (e.g.\ Figure \ref{fig:7465lines}), and then summing the spectra over velocity ranges defined by the \tweco\ emission.  This procedure ensures that the same data volume is used for both lines of a ratio, even if their signal-to-noise ratios are very different.  Spatially-resolved line ratios measured in this manner are presented in Figures \ref{7465radial1} and \ref{7465radial2}.
Reference column densities for the nuclear spectrum in the optically thin and local thermodynamic equilibrium (LTE) approximations are listed in Table \ref{tab:coldens} and line fluxes for all of our defined regions are presented in Appendix \ref{app:bigtab}.

For an alternate method of exploring spatial variations in line ratios, we also implemented a ``shift and stack" technique.  In this method all of the spectra within a given annulus (about the nuclear 3mm continuum peak) can be aligned in velocity, using the \tweco\ velocity field at each position to determine the velocity shift.  Thus all of the expected signal can be concentrated in a narrow range of frequencies.  However, because of the departures from azimuthal symmetry in the center of NGC~7465, ratios produced by this technique have significantly larger uncertainties than those derived from individual regions.

\begin{deluxetable}{llrr}
\tablecaption{Line fluxes and reference column densities in the nucleus of NGC~7465\label{tab:coldens}}
\tablehead{
\colhead{Species} & \colhead{Line flux} & \colhead{$\log N$ (20 K)} & \colhead{$\log N$ (90 K)} \\
\colhead{}        & \colhead{(\jykms)} & \colhead{(cm$^{-2}$)} & \colhead{(cm$^{-2}$)}
}
\decimals
\startdata
\tweco & \phn \phn 3.52 (0.04) & 17.14 &  17.69 \\
\thirco & \phn \phn 0.096 (0.014) & 15.65 & 16.20  \\
\ceighto & $<$ 0.039 &  $<$ 15.26 & 15.81 \\
HCN & \phn \phn 0.132 (0.012)  &  12.80 &13.37  \\
\hcop &  \phn \phn 0.303 (0.013) & 13.39 & 13.96 \\
CN &  \phn \phn0.226 (0.006) & 13.61 & 14.16 \\
CS &  \phn \phn 0.041 (0.009) & 13.02 & 13.54 \\
\cch &  \phn \phn0.091 (0.014) & 14.70 &  15.27 \\
\chhhoh & $<$ 0.035 & $<$ 13.95  &  $<$ 14.57 \\
SiO  & $<$ 0.040 & $<$ 12.79  &  $<$ 13.33 \\
HNCO & $<$ 0.037 & $<$ 13.49 & $<$ 14.19 \\
\enddata
\tablecomments{Line fluxes are measured in a circular region 2\arcsec\ (280 pc) in diameter, centered on the nuclear 3mm continuum source; column densities are averaged over this region.  Statistical uncertainties of the line fluxes are listed in brackets, or the upper limits are quoted as 3 times the statistical uncertainty of the sum over the adopted velocity range.  
For \thirco\ the peak intensity is offset from the nucleus and a 2\arcsec\ region centered on the peak has a line flux of 0.015 \jykms.
Column densities are calculated for two representative temperatures in the optically thin LTE assumption.
Current data do not constrain the optical depths of the HCN and \hcop\ transitions, but CN and \cch\ are optically thin (Section \ref{sec:cch}) and even \tweco\ might be at this position (Section \ref{sec:disc_isotop}).
}
\end{deluxetable}

\begin{figure}
\includegraphics[width=\columnwidth, trim=5mm 5mm 1cm 5mm]{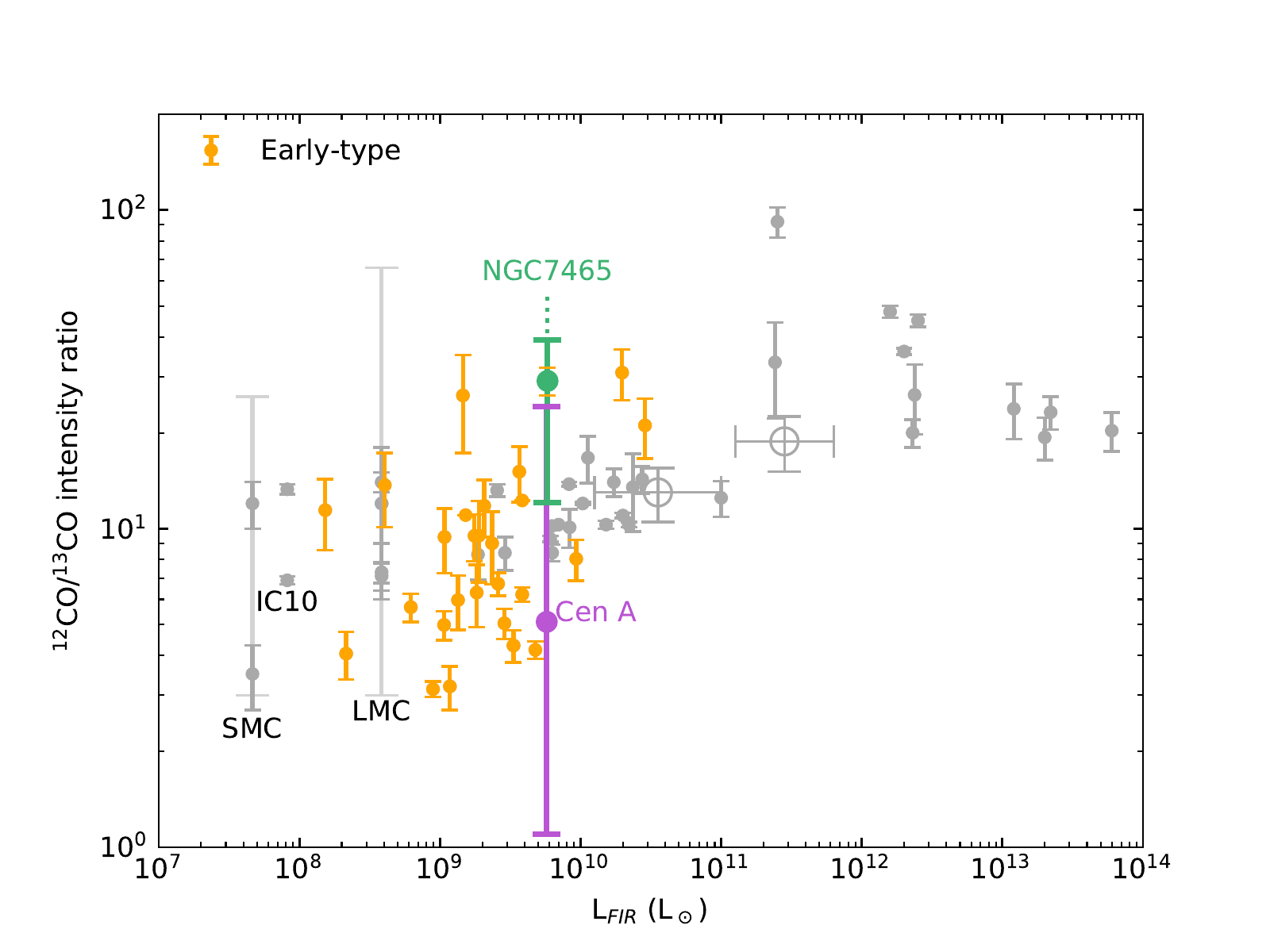}
\caption{A compilation of observed \tweco/\thirco\ line ratios, in the same spirit as Figure 1 of \citet{zhang2018}.  Symbols in grey are dwarf galaxies, spirals, and ULIRGs; at the highest luminosities they are sub-mm galaxies at $z \approx 2-3.$  Data are drawn from \citet{davis_13co}, \citet{jd_13co}, \citet{brown+wilson}, \citet{cormier18}, \citet{henkel14}, \citet{braine2017}, and \citet{heikkila99}; see also the citations in \citet{zhang2018}.  
The large light error bars for the SMC and the LMC show the full range of ratios measured by \citet{israel03} at 12 pc resolution. Open grey circles are stacks of multiple galaxies \citep{mendez-hernandez}.
We use an improved FIR luminosity for IC~10 from \citet{remy-ruyer15}.
Line ratios of early-type galaxies (orange symbols) are compiled from \citet{crocker_hd} and \citet{a3d_13co}; Cen~A is from \citet{mccoy_cena}.
For NGC~7465 and Cen~A (green and purple, respectively), the error bars show the full range of ratios measured in the galaxy.
\label{fig:lir2}}
\end{figure}

NGC~7465 has relatively weak \thirco\ emission.
Measured \tweco(1-0)/\thirco(1-0) ratios are highest at 39 $\pm$ 9 in the nucleus, or 33 $\pm$ 5 when averaged over a circle of 1\arcsec\ radius, decreasing to  $\approx$ 15 $\pm$ 2 at a radius of 1 kpc (see Figures \ref{fig:7465_1213} and \ref{7465radial1}).\footnote{All line ratios reported in this paper are computed using integrated fluxes expressed in temperature units (K \kms).}
The large ratios in the center of NGC~7465 are higher than those typically found in nearby spirals, which tend to be in the range of 8--20 \citep[especially on scales $>$ 1 kpc, e.g.][]{crocker_hd,carmasting,cormier18,israel2020}.  For comparison, Figure \ref{fig:lir2} shows typical \tweco/\thirco\ line ratios for galaxies of various types and illustrates that the
high ratios in the nucleus of NGC~7465 are similar to those found in ULIRGs and advanced mergers such as Arp 220 and NGC~2623 \citep{brown+wilson}.  Figure \ref{fig:lir2} also shows that the range of \tweco/\thirco\ line ratios exhibited by early-type galaxies is a factor of 10, broader than the range exhibited by any other galaxy type.  Other early-type galaxies known to have high \tweco/\thirco\ ratios include NGC~1266 and UGC~09519 \citep{crocker_hd}; NGC~1266 is remarkable for having a strong AGN-powered molecular outflow \citep{ngc1266}, while UGC~09519 has a kinematically-misaligned \hi\ disk that suggests it was recently acquired \citep{serra14}.  NGC~7465 is thus consistent with the pattern that high \tweco/\thirco\ ratios tend to be associated with major disturbances to the ISM.

NGC~7465's strong radial gradient in \tweco/\thirco\ is also uncommon.  It amounts to a factor of two over an unresolved length scale, such that the ratios at radii $<$ 140 pc are at least a factor of two higher than those outside and possibly more.
For context, the spirals in \citet{paglione01}, \citet{carmasting}, and \citet{cormier18} sometimes have enhancements of \thirco\ in their nuclei and sometimes deficits, but none of them are as extreme as in NGC~7465.
Two barred lenticular galaxies are in fact similar to spirals in this respect \citep{topal}, with typical \tweco/\thirco\ ratios about 5 to 15, and with the higher ratios in the central kpc or so; they only exhibit modest gradients of a factor of two over kpc scales.
Measurements at sub-kpc resolution sometimes show spatial gradients of the same magnitude as those in NGC~7465.
\citet{meier2004} found variations of factors of 5 over $\approx$ 200 pc in the center of NGC~6946, but unlike NGC~7465, the spiral shows an enhancement of \thirco\ in its center.
The strong deficit of \thirco\  in the center of NGC~7465 is more reminiscent of the nearby early-type galaxy Cen A, where \citet{mccoy_cena} find \tweco/\thirco\ $>$ 20 in the circumnuclear disk at radii $<$ 200 pc and \tweco/\thirco\ reaching nearly to unity farther out in the arms.

\begin{figure}
\includegraphics[width=\columnwidth,trim=1cm 3mm 1cm 1cm,clip]{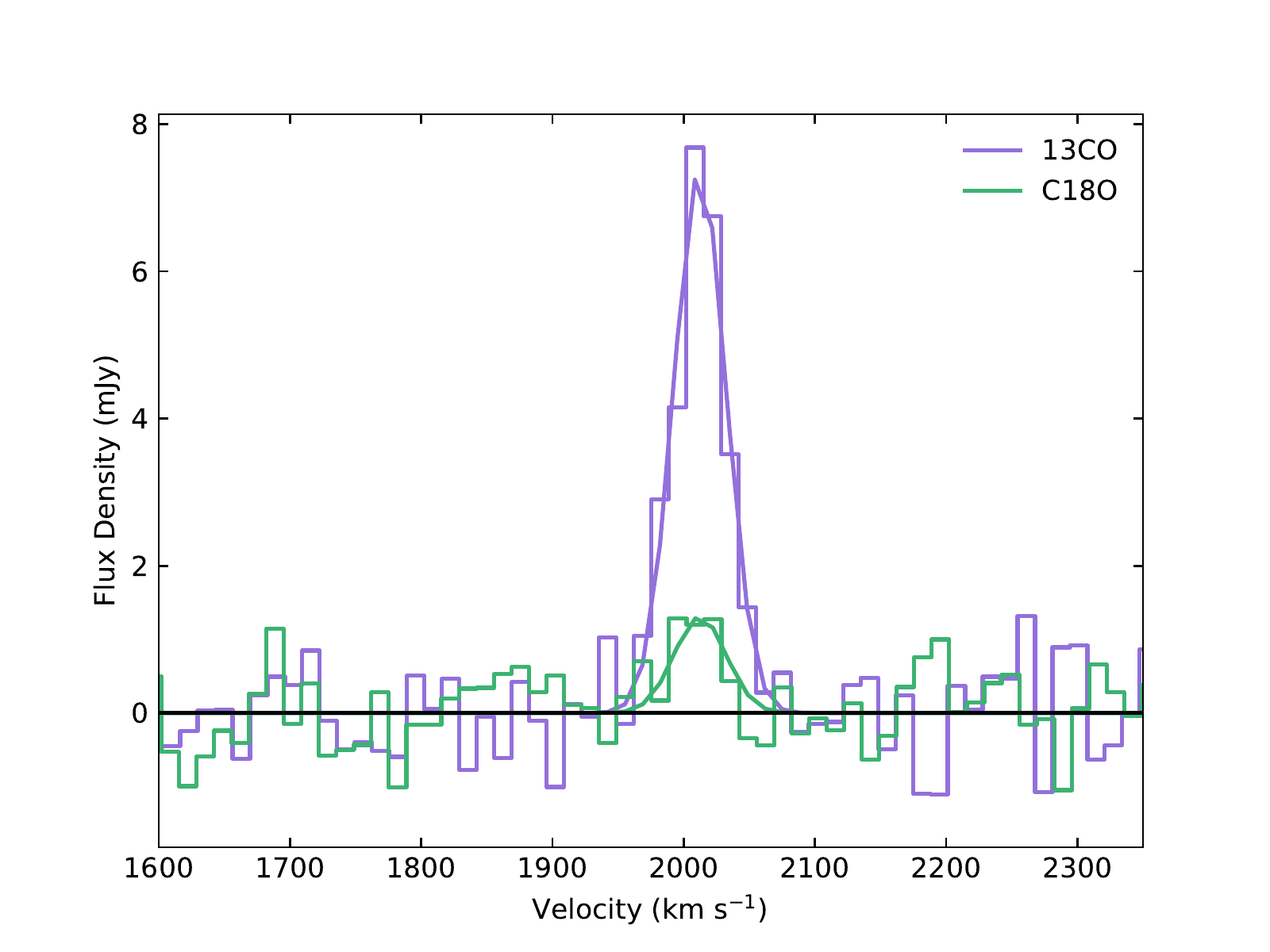}
\caption{\thirco\ and \ceighto\ in NGC~7465.   These spectra are from a region at the northeast end of the molecular ridge, where the \thirco\ emission is brightest.   Gaussian fits are overlaid, though our analysis uses line ratios derived from simple sums rather than from the fits. Since the \ceighto\ line is so faint, its width and center velocity have been constrained to match those of \thirco\ in the fits. \label{7465_1318spec}}
\end{figure}

\begin{figure}
\includegraphics[width=\columnwidth, trim=5mm 5mm 1cm 5mm]{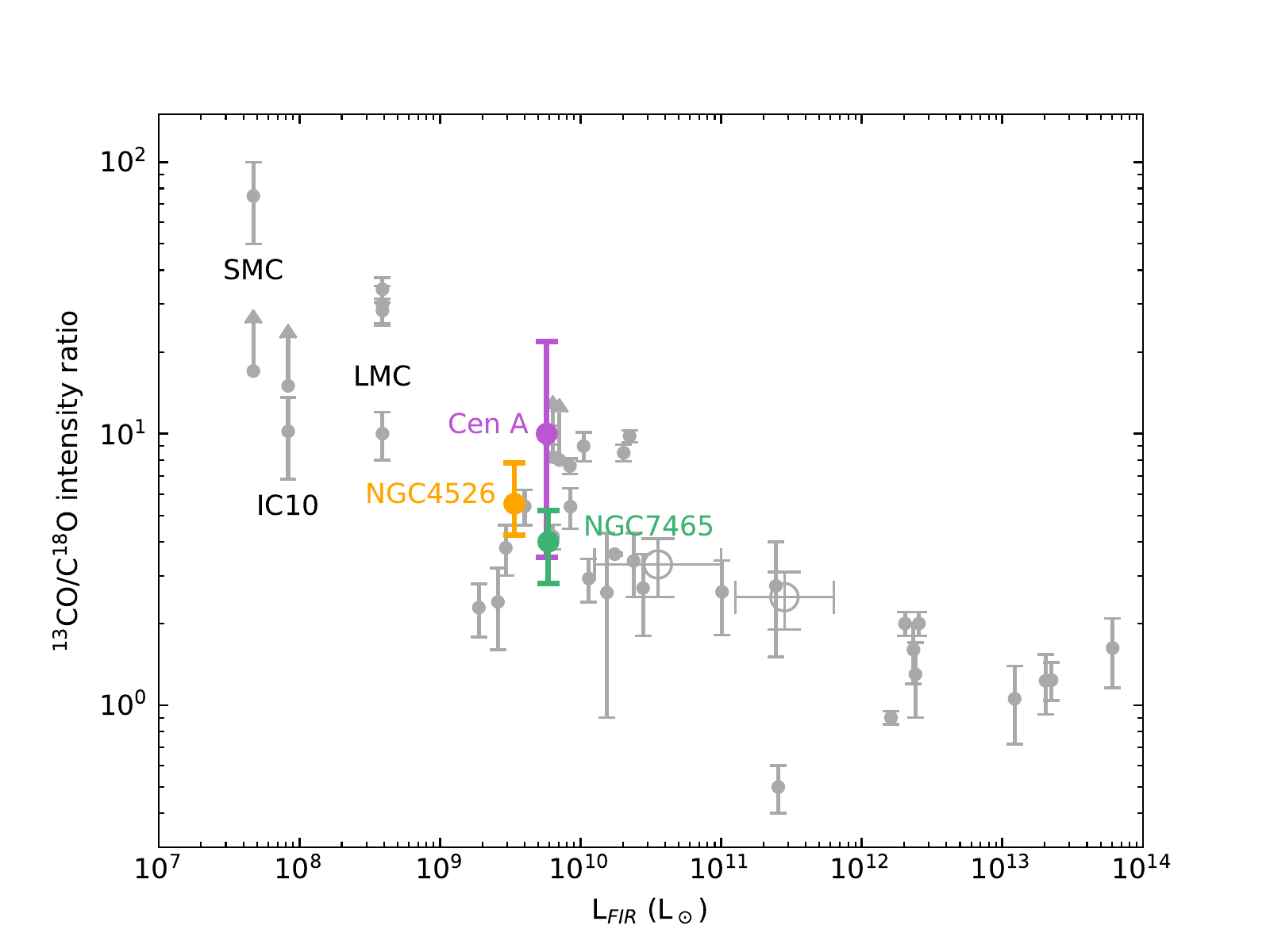}
\caption{A compilation of observed \thirco/\ceighto\ line ratios, after Figure 1 of \citet{zhang2018}.  For NGC~4526 (Young et al.\ in prep) and Cen~A \citep{mccoy_cena}, the error bars show the full range of ratios measured in the galaxy, with the symbol at the median ratio.  In addition to the citations noted in Figure \ref{fig:lir2} and \citet{zhang2018}, this figure includes measurements by \citet{johansson94} and \citet{wang09}.
\label{fig:lir}}
\end{figure}

Measurements of \thirco/\ceighto\ in NGC~7465 are limited to a few spatial regions with detections of \ceighto\ (e.g.\ Figure \ref{7465_1318spec}).  In a large spatial region defined by all of the detected \tweco\ emission, we have a tentative detection of \ceighto\ yielding \thirco/\ceighto\ = 3.0 $\pm$ 1.0.  The bright molecular ridge has a ratio of 4.0 $\pm$ 1.2.
Figure \ref{fig:lir} shows that these ratios are similar to those of galaxies of comparable IR luminosity \citep[e.g.][]{jd_13co}.
Further interpretation of the CO isotopologue line ratios is in Section \ref{sec:disc}.

\section{High-Density tracers}

Figures \ref{7465radial1} and \ref{7465radial2} show that, as expected, the emission from HCN, \hcop, and \cch\ is enhanced relative to \tweco\ and \thirco\ in the center of NGC~7465.  This pattern is common in nearby galaxies and is expected to be driven by higher gas densities in the centers \citep[e.g.][]{jd_bigsurvey}.  We also observe radial trends in \cch/HCN and \cch/\hcop, in the sense that \cch\ is more strongly enhanced just outside of the nucleus than in the nucleus itself.
HCN/\hcop\ and \tweco/CS are effectively constant wherever we can measure them, but CS/HCN increases with radius.  All of these results are discussed in greater detail below.

\subsection{HCN and \hcop}

Relative to \tweco, NGC~7465 has fairly typical HCN brightness compared to other nearby galaxies.  
Specifically, \tweco/HCN in Figure \ref{7465radial1} ranges from 16 $\pm$ 1 to 50 $\pm$ 15.
These ratios are very similar to their analogs measured in the nearby spirals of the EMPIRE survey by \citet{jd_bigsurvey} and in two other lenticulars by \citet{topal}.
The EMPIRE data have somewhat lower linear resolution (1 to 2 kpc) than our measurements in NGC~7465 (240 pc) but in all of these cases the disks are well resolved.

In contrast, the HCN in NGC~7465 is relatively faint compared to that in a larger sample of early-type galaxies;  in the unresolved single-dish measurements of \citet{crocker_hd}, NGC~7465 has the highest \tweco/HCN ratio of the sample.  The unresolved measurements undoubtedly fold in some radial gradients, so it is difficult to make detailed comparisons at different scales and it would be worthwhile to obtain additional resolved \tweco/HCN measurements in early-type galaxies.

The large difference in critical or effective density between CO and HCN implies that observed ratios like \tweco/HCN encode information about the density distribution of the molecular gas, or equivalently the relative proportions of low-density and high-density gas \citep[e.g.\ $10^2$ -- $10^3$ \percc\ vs.\ $10^5$ -- 10$^6$ \percc;][]{meier2012, leroy2017}.  Thus, NGC~7465 has a mix of low/high-density gas which is typical of most nearby spirals, but it is relatively deficient in high-density gas compared to many other local early-type galaxies.  The proportion of high-density gas is maximized in the nucleus but is also high towards the southern continuum source, where we find the second-lowest \tweco/HCN and \tweco/\hcop\  ratios in the galaxy as well as some of the lowest reliably-measured \tweco/\thirco\ ratios.  This region must contain a concentration of dense gas with relatively high CO optical depth.

Relative to \tweco, NGC~7465 is exceptionally bright in \hcop\ emission when compared to nearby galaxies. 
We measure local \tweco/\hcop\ ratios from  $6.9 \pm 0.3$ in the nucleus to $20.7 \pm 2.9$ in the disk.  
In comparison, the spiral galaxies of the EMPIRE survey usually have \tweco/\hcop\ $> 25$ even in their nuclei \citep{jd_bigsurvey}. 
Thus, the regions of NGC~7465 with the {\it faintest} \hcop, at radii $\approx 1$ kpc, still have {\it brighter} \hcop\ than is common even in the nuclei of the spirals.
It is also notable that the \hcop\ emission in the nucleus of NGC~7465 has a dramatically different line shape than \tweco\ (Figure \ref{fig:7465lines}), which probably reflects differing spatial distributions.

\subsection{HCN/\hcop}

Because \hcop\ is so bright, the HCN/\hcop\ ratios in NGC~7465 are low; they are significantly $< 1$ everywhere we are able to measure them, and this is unusual for nearby galaxies.
We measure HCN/\hcop\ = 0.44 $\pm$ 0.04 in the nucleus, with ratios elsewhere ranging up to 0.60 $\pm$ 0.11.
In nearby galaxies it is much more common to find HCN/\hcop\ $>$ 1 \citep[e.g.][]{jd_bigsurvey,topal}.
Ratios $<$ 1 are seen in restricted parts of a few spirals \citep[e.g.\ parts of NGC~4254 and Maffei 2;][]{meier2012,gallagher2018}.
Low ratios of HCN/\hcop\ $\approx$ 0.5 are occasionally seen in other early-type galaxies \citep{crocker_hd} and more commonly in
ULIRGs and starburst galaxies \citep[e.g.][]{baan2008,krips08,privon2015,sliwa+downes}.
Extremely low HCN/\hcop\ ratios $\lesssim$ 0.2 have been measured in dwarf galaxies like the LMC and IC~10 \citep{seale2012,braine2017,kepley2018,anderson_lmc}.  In this respect, NGC~7465 has more in common with starburst or dwarf galaxies than with nearby spirals. 
The low HCN/\hcop\ and HCN/CO ratios in NGC~7465 also place it firmly amongst the nuclear starburst galaxies and ``composite" Seyferts in the HCN(1-0) diagnostic diagram of \citet{kohnohcn}.

Spatial variations in the HCN/\hcop\ ratios of NGC~7465 do not significantly exceed the ratios' mutual statistical uncertainties, so there is no compelling evidence for a radial gradient (Figure \ref{7465radial2}).
Even though there is an AGN in NGC~7465, there is definitely no enhancement in HCN/\hcop\ towards the nucleus, 
unless it is on length scales of tens of pc where it would be obscured by our spatial resolution.
There is certainly nothing like the factor of 2 rise in HCN/\hcop\ seen in the central 0.5 kpc of NGC~1068 \citep{viti2014}.

\subsection{Physical properties of the dense molecular gas}

Due to the difference in critical densities, the HCN/\hcop\ line ratio is most commonly interpreted as an indicator of density in the high-density ($n \sim 10^4 - 10^6$ \percc) portion of the molecular gas.
However, for the lowest HCN/\hcop\ line ratios in the LMC, \citet{anderson_lmc}
have appealed to a combination of low density and low gas-phase metallicity, with low metallicity contributing in two ways.  First, lower heavy element abundances will produce lower column densities of these species and smaller optical depths in their transitions, which is necessary for the line ratio to deviate from 1.  On top of that,  lower metallicities are associated with lower [N/O] abundance ratios \citep{vanzee06},
 which might manifest as lower abundances of HCN relative to \hcop\ \citep{braine2017}.

NGC~7465 does not have a low metallicity (Section \ref{cont}), so there is no reason to assume a low [N/O] ratio {\it a priori}.  However, the observed line ratios in NGC~7465 do constrain the abundance ratio [HCN/\hcop] to be significantly smaller than that inferred for NGC~253 by \citet{meier2015}, which was [HCN/\hcop] $\approx$ 5.
In LTE, reproducing our observed line ratios of HCN/\hcop\ = 0.44 $\pm$  0.04 requires an abundance ratio [HCN/\hcop] $\leq 0.25$ (Table \ref{tab:coldens}).\footnote{We adopt the convention that ratios written without brackets (e.g.\ HCN/\hcop) refer to measured line ratios whereas ratios inside square brackets (e.g.\ [HCN/\hcop]) refer to estimated molecular abundances after correcting for optical depth and excitation effects.}
In the more likely situation that LTE does not apply, we make calculations using the RADEX large velocity gradient code \citep{radex} to account for optical depth and excitation effects on the HCN/\hcop\ line ratio; broad ranges of parameters will reproduce the data from NGC~7465 but they always require an abundance ratio [HCN/\hcop] $\lesssim 3,$  and a ratio of 5 is ruled out as it cannot reproduce the line ratios in NGC~7465.
For an assumed [HCN/\hcop] = 3, relatively low densities $n_{\rm H_2} \leq 10^{4.3}$\percc\ are required to reproduce the observed line ratios.
Thus, even though the metallicity of NGC~7465 is not low enough to suggest unusually low N abundance, the observed HCN/\hcop\ line ratios in NGC~7465 are best explained with a combination of relatively low-density gas and/or a relatively low HCN abundance (or enhanced \hcop).  More specific estimates of the density and temperature will require additional transitions or isotopologues, but the inference of a low HCN abundance is robust to other variations in physical properties.

Viewing the molecular gas as two ``phases" -- a more diffuse phase with densities $\sim 10^2$ to $10^3$\percc\ traced by CO, and a denser phase $\gtrsim 10^5$\percc\ traced by HCN and similar species -- the strong spatial trends in diffuse/dense gas tracers suggest that the relative proportions of the two phases change with radius in NGC~7465.  However, the internal properties of the denser phase do not seem to change, even though the properties in the more diffuse phase are dramatically varying (Section \ref{sec:disc_isotop}).

\section{\cch\ and CN: photodissociation regions}\label{sec:cch}

The \cch\ molecule is regarded as a tracer of photon-dominated or photodissociation regions (PDRs), as its formation is driven by the presence of C$^+$, and it has been shown to be brighter on the illuminated sides of molecular clouds 
\citep[e.g.][]{meier2005,meier2012,meier2015,garcia_cch}.
The two main fine structure blends in our data are $N=1-0$, $J=\frac{3}{2}-\frac{1}{2}$, with a rest frequency of 87.3169 GHz,  and $N=1-0$, $J=\frac{1}{2}-\frac{1}{2}$, at 87.40199 GHz.  They are separated by 292 \kms, which is just greater than the velocity range covered by \tweco\ in NGC~7465, so they are not overlapping.  
The low-frequency component is detected in NGC~7465 but the 
high-frequency component is not; as the high-frequency component is a factor of 2.3 fainter in LTE in the optically thin limit, its nondetection is not surprising, but we can rule out line ratios $\approx$ 1 (the optically thick limit).  Thus we cannot measure the \cch\ optical depth directly but the data are consistent with the transitions being optically thin.  Further, the \cch\ line fluxes quoted in this paper refer only to the blend at 87.32 GHz.  

The  \cch\ emission in NGC~7465 is relatively bright, when compared to HCN.
We find \cch/HCN ratios ranging from 0.71 $\pm$ 0.13 in the nucleus to 1.0 $\pm$ 0.2 in the outer parts of the molecular ridge.
These are significantly higher than the corresponding ratios in the central kpc of NGC~253, which range from 0.17 to 0.47 \citep[$\pm$ 15\%;][]{meier2015}.  The \cch/HCN ratios in NGC~7465 are also on the high side when compared to the set compiled by \citet{martin_cch} for active and starbursting galaxies.
On the other hand, ratios of \cch/\hcop\ in NGC~7465 are very similar to those in NGC~253 and other galaxies.  We find 0.31 $\pm$ 0.05 to 0.50 $\pm$ 0.11 in NGC~7465, where Meier et al.\ find 0.20 to 0.57 ($\pm$ 15\%) in NGC~253 and most of the sample galaxies in \citet{martin_cch} show a similar range.
These results suggest that HCN emission in NGC~7465 is relatively faint because of moderate densities in the molecular gas; both \cch\ and \hcop\ are relatively prominent because their critical densities are a factor of 10 lower than that of HCN.

In addition to being bright, the \cch\ emission in NGC~7465 is also notable in that 
the radial trends of \cch/HCN and \cch/\hcop\ in NGC~7465 are opposite to the trends in NGC~253.  In the spiral those ratios peak at the nucleus; in other words, \cch\ is more strongly concentrated than HCN and \hcop\ are.
In NGC~7465 it is the opposite -- \cch\ is less centrally concentrated than HCN and \hcop.  The radial trend in NGC~7465 thus suggests
enhanced \cch\ emission just outside of the nucleus, at radii 2\arcsec\ to 5\arcsec\ or 280 to 700 pc (as far out as we can detect \cch).
If the C$^+$ ionization is indeed dominated by star formation activity, this radial trend
might be consistent with the ionized gas emission line ratios in Figure \ref{fig:7465cont} (Section \ref{cont}): the lowest \oiiihb\ ratios (most suggestive of star formation activity) occur slightly outside the nucleus at radii $\approx$ 3\arcsec\ to 10\arcsec.
Similarly, from the stellar population analysis of \citet{davor_ageZ}, there is a suggestion that the smallest stellar ages occur at a radius of 2.\arcsec5 rather than at larger or smaller radii.

As another tracer of C$^+$, we note that NGC~7465 is also detected in \cii\ 158\micron\ emission \citep{lapham2017}, though the spatial resolution is not good enough to compare to the radial trends in our ALMA data.
Many spiral galaxies are known to have significant central deficits of \cii\ emission at radii $<$ 1 kpc, and many potential explanations have been proposed \citep[e.g.][]{jdsmith_cii}.  Possibly, in some cases, the central \cii\ deficit arises because an AGN's hard radiation field drives the ionization balance towards C$^{2+}$;  in other cases, a softer radiation field (caused by old stellar populations) may be driving the ionization balance towards neutral carbon.  At present we simply comment that the central \cch\ deficit in NGC~7465 (relative to HCN and \hcop) might be consistent with what is known about the ionization states of C in other galaxy nuclei.

The CN(1-0) transition is also a PDR tracer \citep{boger_cn}, though it is expected to be even more strongly enhanced in X-ray dominated regions (XDRs) and may thus help to distinguish between PDRs and XDRs \citep{meijerink2007}.
It has fine structure components that appear in these data as blends with rest frequencies of about 113.17 GHz and 113.49 GHz; they are clearly separated in our data and both are clearly detected.  The high-frequency blend is the brighter one and there is no evidence that their line ratio in NGC~7465 ever deviates from the theoretical optically thin limit of 2 \citep[e.g.][]{meier2015}.  Thus, the CN emission in NGC~7465 is optically thin.  For simplicity, we show only the brighter component in the figures and the line fluxes we quote refer only to it.

The range of \tweco/CN ratios we measure in NGC~7465 is large, from $15.0 \pm 0.4$ in the nucleus to $93 \pm 14$ when averaged over the full \tweco\ emission region, and it is consistent with the wide range of ratios measured in other nearby galaxies \citep{wilson_cn}.  CN/HCN ratios are consistent with 1 aside from one region, 
and this again is very typical of local galaxies \citep{ueda_cn,cicone_cn}.

There is some evidence that CN is even more centrally concentrated in NGC~7465 than the other high-density tracers like HCN.  The radial variations of \tweco/CN exceed those of all other \tweco\ ratios (Figure \ref{7465radial1}).   The transition in line profile shapes, from double-peaked \tweco\ through flat-topped \hcop\ to centrally-peaked CN (Figures \ref{fig:7465lines} and \ref{fig:7465pvs}), might thus suggest the presence of a high-density circumnuclear disk at $r \lesssim 100$ pc.  CN might be particularly enhanced close to the AGN because of its X-ray sensitivity \citep{meijerink2007}.  Higher resolution data, especially for CN, would be valuable for testing this scenario.  As a caveat we note that the current data do not show a measurable radial trend in CN/HCN (Figure \ref{7465radial2}), but such a trend could be obscured by the resolution and sensitivity of our data.
We do find a systematic difference in the spatial distribution of the two PDR tracers, \cch\ and CN; \cch\ is less centrally concentrated than HCN whereas CN is equally or possibly more centrally concentrated than HCN.  The differences between these two species might be consistent with an interpretation that \cch\ is a better tracer of star formation activity whereas CN is a better tracer of XDRs. 

\section{CS}\label{sec:cs}

The CS molecule is also a high-density tracer, whose $J=2-1$ transition has a critical or effective density intermediate between those of HCN and \hcop\ $J=1-0$ \citep[e.g.][]{leroy2017}.
There is also some evidence that CS emission is enhanced in PDRs, where the radiation field is stronger \citep{lintott_cs,meier2005}, which has motivated suggestions that CS emission traces massive star formation even better than HCN emission does \citep[e.g.][]{bayet09,davis_cs}.

NGC~7465 has \tweco/CS ratios that are fairly typical compared to those of other nearby galaxies.
We find \tweco/CS = 62 $\pm$ 13 in the nucleus, with variations of no more than 25\% in all of the regions where we detect CS.
For comparison, the spirals in \citet{gallagher2018} have typical ratios of \tweco/CS $\approx$ 90 in their central kpc, so NGC~7465 is brighter in CS than these.
However, it is faint in CS compared to the other early-type galaxies in \citet{davis_cs}, which have \tweco/CS in the range of 9 to 44.
The spatially-unresolved (IRAM 30m) ratios in \citet{davis_cs} might also be biased high if the beam-filling factor of \tweco\ is higher than CS, so that resolved measurements of CS would probably increase the discrepancy between them and NGC~7465.

CS/HCN, which should provide better insight (than \tweco/CS) into the properties of the dense molecular phase, is very similar in NGC~7465 to other nearby galaxies.
We find 0.26 $\pm$ 0.06 in the nucleus of NGC~7465 and 0.57 $\pm$ 0.14 for the larger regions that extend to 0.8 kpc.  These ratios are consistent with those in the spirals Maffei 2 and NGC~253 \citep[0.2 to 1;][]{meier2012,meier2015}.  The spirals in \citet{gallagher2018} also typically have CS/HCN $\approx$ 0.1 to 0.3, with localized ratios up to 0.5 in the central 200 pc of NGC~3627 or the arms of NGC~4321.  
CS/HCN in NGC~7465 is also consistent with the single-dish measurements of some other early-type galaxies in \citet{davis_cs}.
It is notable that others of the early-type galaxies in that sample have highly unusual ratios of CS/HCN as large as 3, so that bright CS emission might be a feature of some early-type galaxies.

In terms of radial trends, it is striking that the nucleus of NGC~7465 does not differ from the off-nuclear regions in terms of \tweco/CS; in contrast, there are strong radial trends of factors of 2--4 in \tweco/HCN, \tweco/\hcop, \tweco/\cch\ and \tweco/CN.
We also find no radial variation of HCN/\hcop\ but a factor of two in CS/HCN.
If the density of the molecular gas were the only factor driving changes in those line ratios, we would expect HCN, \hcop, and CS to behave in qualitatively the same fashion as the critical density of CS is intermediate between those of \hcop\ and HCN.  Evidently, then, we require more than simple density variations to explain the behavior of these high-density tracers in NGC~7465.

\citet{davis_cs} have also studied unresolved CS emission in a small sample of early-type galaxies.  They compared it to \oiiihb, which can serve as an indicator of whether the ionization in the optically-emitting ionized gas is dominated by AGN activity or star formation (see also Section \ref{cont}).  In their early-type galaxies, the global \oiiihb\ ratio is correlated with CS/HCN in the sense that galaxies whose ionization is more dominated by star formation (smaller \oiiihb; softer radiation fields) have higher CS/HCN.
NGC~7465 is consistent with this trend, both in terms of its integrated line ratios and its internal spatially-resolved behavior, for which we find higher CS/HCN and lower \oiiihb\ (and bluer optical colors) at $r \approx$ 0.5 to 1 kpc than towards the nucleus.
NGC~7465 thus provides circumstantial evidence supporting the interpretation that CS emission is enhanced by PDR conditions.

For the ionized gas metallicity estimates in Section \ref{cont} and the column density estimates in Table \ref{tab:coldens}, NGC~7465 is also consistent with the suggestions of \citet{bayet12} and \citet{davis_cs} that CS/HCN could be used as an indicator of the metallicity of the molecular gas.  Future work on spatially-resolved nebular emission line data might also be valuable to test whether the trend in \citet{davis_cs} also applies within individual galaxies.

\section{Shock tracers}

Emission from SiO, \chhhoh, and HNCO is commonly interpreted as tracing shocks in the ISM \citep[e.g.][]{meier2012,meier2015}, as significant energy input in the form of shocks is required to liberate them from grain surfaces.
These molecules are not detected in NGC~7465, with the most stringent limits being \tweco/(SiO, \chhhoh, or HNCO) $>$ 83 near the center of NGC~7465.  
Ratios relative to \tweco\ are not particularly significant in terms of a physical interpretation, but here merely serve as an indication of the relative faintness of the lines compared to detections in other galaxies.
For context, in NGC~253, \citet{meier2015} found spatially-resolved ratios of \tweco/SiO and \tweco/HNCO $\approx$ 20 to 80;  NGC~253 thus has brighter SiO and HNCO than NGC~7465, by at least a factor of 4 relative to \tweco.
The circumnuclear disk of NGC~1068 has \tweco/SiO = 12.5 $\pm$ 1.5 \citep{garcia2010}, which again is at least a factor of 7 brighter in SiO than NGC~7465.
Finally, \citet{topal} also measured \tweco/HNCO = 66 $\pm$ 5 and 78 $\pm$ 18 in the centers of the lenticulars NGC~4710 and NGC~5866.  Thus, these two also have modestly brighter HNCO than NGC~7465.

For \chhhoh, \citet{davis_cs} detected a small sample of early-type galaxies with unresolved measurements of \tweco/\chhhoh\ in the range 14 to 70.  As we measure \tweco/\chhhoh\ $>$ 83, we conclude that NGC~7465 is also relatively faint in \chhhoh\ emission.
Similarly, unresolved measurements of \chhhoh/HCN reach as high as 1.5 to 2.2 in the early-type galaxies NGC~6014 and NGC~5866 \citep{davis_cs}, whereas the resolved upper limits in NGC~7465 are 0.2 to 0.8.

In short, SiO, \chhhoh, and HNCO  are relatively weak in NGC~7465 compared to other galaxies that have been studied to date.
The relatively weak emission from these shock tracers is a bit surprising, given the obvious disturbances in the galaxy (Sections \ref{why7465} and \ref{gasdist}).  The molecular and \hi\ disks are strongly warped and their kinematics are significantly misaligned with respect to those of the stars, so it is clear that the gas was recently acquired from an external source.  We might have expected stronger emission from shock tracers as the gas disk is still in the process of settling; possibly the shocks are primarily occurring farther out in the galaxy, where our sensitivity is not currently good enough to detect the shock tracers.

\section{Discussion: constraints on abundances and galaxy evolution}\label{sec:disc} 

\subsection{CO isotopic ratios in NGC~7465}\label{sec:disc_isotop}

Figure \ref{fig:lir2} shows that the center of NGC~7465 boasts an unsually high \tweco/\thirco\ line ratio; it is reminiscent of the ratios commonly observed in LIRGs and ULIRGs rather than those in spirals and other early-type galaxies.
It also constrains the [\tweco/\thirco] abundance in the nucleus of NGC~7465 to be $\geq$ 39 $\pm$ 9, which is consistent with most measurements in the disk of the Milky Way, in other nearby spirals, and in part of the LMC \citep[e.g.][and references therein]{johansson94,meier2008,romano2019}.
However, it is incompatible with the very low abundance ratios of [\twec/\thirc] = 9 ($\pm$ 2), 21 ($\pm$ 6), and 24 ($\pm$ 1) estimated in the centers of NGC~4945, NGC~253, and the Milky Way \citep{langer1990,tang2019,martin2019}.
Thus, the chemical enrichment pattern in the central molecular gas of NGC~7465 differs from the pattern in these nearby spirals.

The smaller \thirc\ abundance in the center of NGC~7465 is plausibly connected with the clear signs of gas accretion, as the gas currently in the center of NGC~7465 was recently
in a different galaxy and possibly in the outskirts of that galaxy.
For context, the stellar dynamical analysis of \citet{cap:a3dJAM} implies a maximum circular rotation speed of 163 \kms, so the orbital time at the outer edge of the molecular arms in our data (15\arcsec\ or 2.1 kpc)
is 80 Myr.  It may have taken a few orbital times for the gas to make its way inward to the center of the galaxy, and this timescale is comparable to or shorter than the timescale for significant \thirc\ production \citep[e.g.][]{romano2019}.

Low inferred optical depths in \tweco(1-0) (see below) also imply that both \thirco\ and \ceighto\ are optically thin and thus the measurements of [\thirco/\ceighto] in NGC~7465 are 2.98 $\pm$ 0.99 and 4.0 $\pm$ 1.2.  These measurements are made in overlapping regions so they are not independent.

The optical depths of the CO transitions also convey information about the physical properties of the molecular gas, and if the [\tweco/\thirco] abundance ratio in the center of NGC~7465 is in the typical range of 40--60, then the \tweco\ emission is unusually optically thin.  We thus consider several possible explanations for the unusually high \tweco/\thirco\ line ratio in the center of NGC~7465 and its strong radial gradient.
 
{\it (1) Unusual isotopic abundances due to a burst of star formation?}
Away from the nucleus, the \tweco/\thirco\ and \thirco/\ceighto\ line ratios in NGC~7465 are similar to those in nearby spirals.
But a very recent burst of star formation in the center of the galaxy would affect the isotopic abundances, as \thirc\ in particular comes from low-mass stars with long lifetimes and a region undergoing a starburst might therefore be deficient in \thirc.   Indeed, this effect probably contributes to the exceptionally high ratios of \tweco/\thirco\ $\approx$ 90 in the advanced merger NGC~2623 \citep{brown+wilson}. 
In NGC~7465, although there is some current star formation activity in the central kpc (Sections \ref{why7465} and \ref{cont}), the star formation rate is modest compared to what we usually think of as starbursts.  The luminosity-weighted mean stellar ages are younger in the center of the galaxy \citep{davor_ageZ} but neither the ages themselves nor the radial gradient in stellar ages are unusual even for early-type galaxies.  Ultimately a full multi-level isotopic abundance study would be necessary to resolve the question, but with current data the motivation for assuming unusual isotopic abundances in NGC~7465 is not compelling.

\begin{figure}
\includegraphics[width=\columnwidth, trim=5mm 5mm 1cm 5mm]{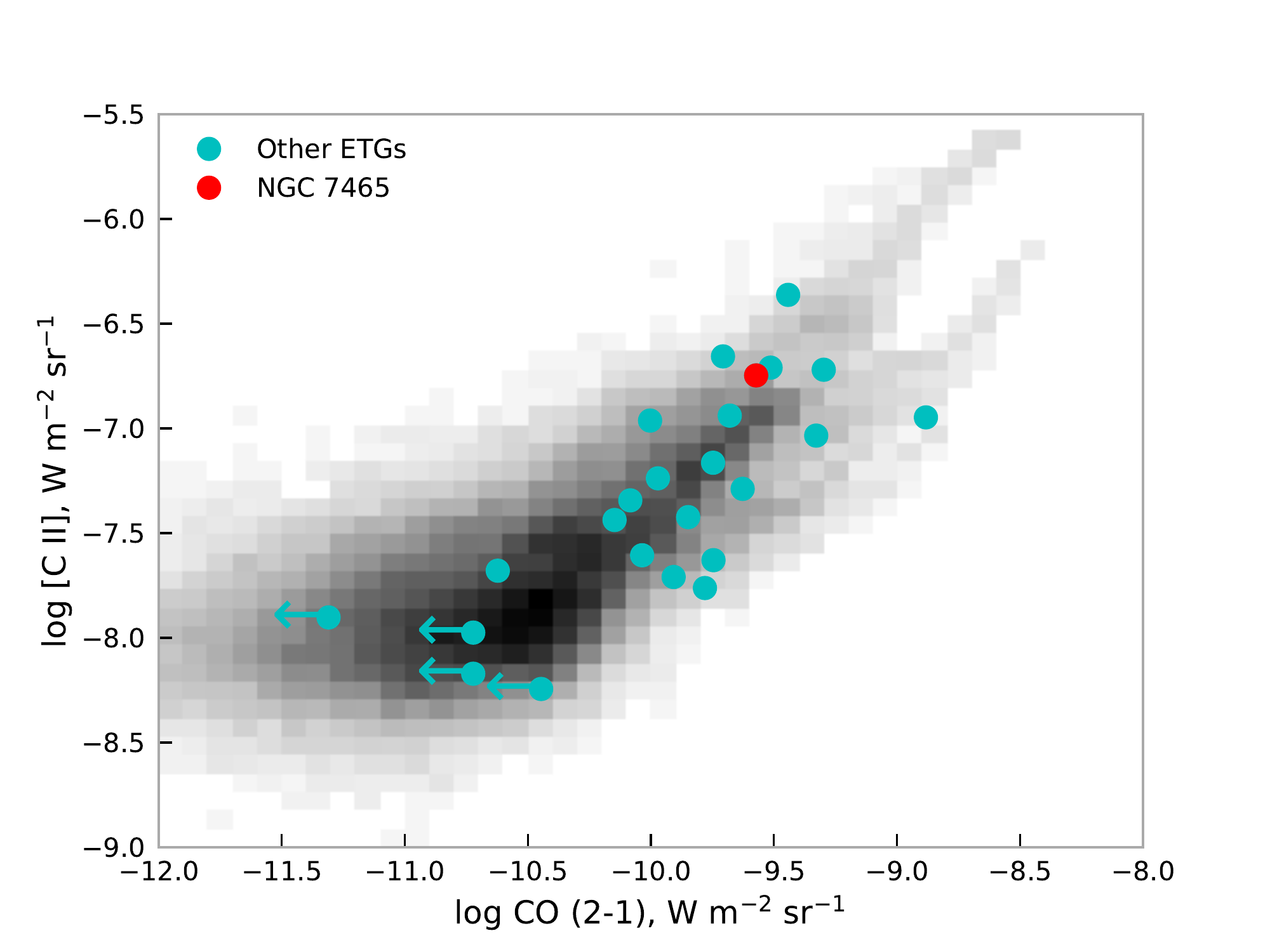}
\caption{Matched-resolution \cii\ and \tweco\ intensities for early-type galaxies \citep[circles;][]{lapham2017, werner, wilson4125, mittal2011, mittal2012}.   For comparison, the greyscale shows a two-dimensional histogram of the same lines in the spiral galaxies of the KINGFISH and HERACLES surveys \citep{kingfish,heracles}.
The angular resolution of all these data is 12\arcsec, which is $\lesssim$ 500 pc in the spirals but $\approx$ 2 kpc in the early-type galaxies.
\label{fig:ciico}}
\end{figure}

{\it (2) Unusual molecular abundances due to fractionation or photodissociation of the rarer isotopes?}
The photodissociation of CO and \htoo\ molecules usually proceeds through absorption-driven excitation, meaning that the molecules can shield themselves from photodissociation if they have sufficient column densities to make the relevant UV transitions optically thick.  \thirco\ and \ceighto\ molecules should then be found deep within the dark interior of a molecular cloud, in the same way (but more so) that \tweco\ molecules should be found deeper than \htoo\ and there may be a skin of CO-dark \htoo\ around a UV-illuminated cloud.
However, this isotope-specific photodissociation is probably not the explanation for the high \tweco/\thirco\ ratio in the center of NGC~7465.  
The \ceighto\ species should be even less abundant than \thirco\ \citep[e.g.][]{meier2008,romano2019}.  Thus, photodissociation effects predict that unusually large \tweco/\thirco\ line ratios should also be accompanied by unusually large \thirco/\ceighto\ ratios,
which we do not find in NGC~7465 (Figure \ref{fig:lir}).  We further note that the center of NGC~7465 does not exhibit particularly unusual levels of photodissociation in general, as its \cii/CO line ratio is typical for nearby spiral galaxies (Figure \ref{fig:ciico}).  Finally, \citet{viti2020} argue that fractionation is unlikely to be a significant effect for the CO isotopologues in typical conditions, though it may be more important for other molecular species.

{\it (3) Variations in the physical properties of the gas?}
The spatial variation of \tweco/\thirco, with higher ratios in the central 100 pc of the galaxy (Figure \ref{fig:7465_1213}), suggests that the optical depth of both molecular species is lower in the center.  For typical abundances and conditions, the optical depth of \tweco\ (1-0) must change by a factor of a few such that it is $>1$ at 1 kpc and $<1$, possibly as low as 0.1, at 100 pc.  Thus, while the \tweco\ column density estimates in Table \ref{tab:coldens} are underestimates, they are not as far off as usually assumed, and they are probably less than a factor of 2 too low.  
Further, the inferred variations of optical depth could be plausibly explained by radial variations of the temperature and/or density of the molecular gas.  

\citet{bayet13} used the $\mathrm{J}=1-0$ and $2-1$ transitions of \tweco\ and \thirco\ to constrain the densities and temperatures of the molecular gas in 18 early-type galaxies, including NGC~7465.  They found the most probable conditions for NGC~7465 to be in the range $10^{4.5}$ to $10^{5.6}$\percc\ and 80 to 150 K.  Notably, these are the highest temperatures inferred for the sample of 18.  These estimates are based on single-dish spectra, so they are spatially unresolved and dominated by the conditions at small radii where the emission is brightest.  However, for these densities and temperatures, calculations using the RADEX code \citep{radex} suggest that increases in the temperature of a factor of 3 (e.g.\ from 30 K to 90 K) could account for the larger \tweco/\thirco\ 1-0 ratio in the central 100 pc of NGC~7465. 
Density changes could have a similar effect on \tweco/\thirco\ but would then also produce a factor of at least 2 variation in HCN/\hcop, which we have ruled out. 
Therefore, unless the CO-emitting gas is completely disconnected from the HCN-emitting gas, temperature variations are more likely than density variations to explain the \tweco/\thirco\ gradient in NGC~7465.
On the other hand, \citet{bayet13} did not make use of resolved kinematic information or the fact that the molecular line widths all increase dramatically towards the nucleus of NGC~7465, which can also affect the line ratios and is discussed in more detail below.

{\it (4) Lower opacity due to higher line width?}
It is striking that the highest \tweco/\thirco\ ratios in NGC~7465 are coincident with the largest linewidths and strongest local velocity gradients in the galaxy (Figure \ref{fig:7465pvs}; Appendix \ref{app:outflow}).  
The \tweco\ line profiles in the disk at radii of 5\arcsec\ to 20\arcsec\ (0.7 to 2.8 kpc) have full widths at half maximum (FWHM) of 23.5 $\pm$ 5 \kms, whereas the line profiles in the nucleus are much broader and double-peaked (Figure \ref{fig:7465lines}).  At 0.\arcsec8 (110 pc) resolution, the line profile toward the nucleus has a FWHM of 230 \kms; that is 10 times larger than in the disk of the galaxy.
And since the optical depth of a transition depends on the ratio of the total column density to the line width \citep[e.g.][]{paglione01,radex}, such a dramatic increase in the line width could produce a corresponding factor of 10 decrease in the optical depth and a rise in the \tweco/\thirco\ ratio.

With the present data it is not possible to distinguish whether the \tweco\ optical depths in NGC~7465 are more significantly affected by temperature changes or increases of the line widths.
Some of the line profiles in the centers of the lenticulars NGC~4710 and NGC~5866 have similarly large linewidths to NGC~7465 but they do not have unusually large \tweco/\thirco\ ratios \citep{topal}.  Thus, it's not yet clear what the difference is between these cases and NGC~7465.
Additional data on higher-energy transitions would help distinguish between the possibilities.

For context, some authors have suggested that mechanical feedback from star formation increases local velocity dispersions in molecular gas, produces low-density, diffuse gas, and contributes to increases in \tweco/\thirco\ ratios near regions of active star formation \citep[e.g.][]{tan2011}.
But in most spirals the variations of linewidths and \tweco/\thirco\ have smaller magnitude than we find.
Conversely, other studies of early-type galaxies have found that dynamically regular, relaxed disks in high-density environments like the Virgo Cluster tend to have low \tweco/\thirco\ ratios, and those could be due to a combination of relaxed kinematics and ram-pressure stripping of low-density gas \citep[e.g.][]{crocker_hd,a3d_13co}.

AGN-driven outflows might also disturb their surrounding molecular gas, increasing its linewidths and its \tweco/\thirco\ line ratios \citep[e.g.\ NGC~1266,][]{crocker_hd}.
Further inspection of the \tweco/\thirco\ ratios in Figure \ref{fig:7465_1213} shows that the ratios are high (the optical depths are low) not just towards the AGN but also in a band stretching along the minor axis of the molecular ridge.  In combination with the unusually misaligned gas kinematics in the center of the galaxy (Figure \ref{fig:7465mom1}; Appendix \ref{app:outflow}), this pattern in the \tweco/\thirco\ ratio image suggests that there might be a modest-velocity ionized gas outflow from the AGN stirring up the gas along the minor axis of the molecular ridge.

\subsection{Implications of the CO isotopic abundances for galaxy evolution}

Comparisons of the stellar and gas kinematics in early-type galaxies provide clear evidence that much -- perhaps as much as half -- of the gas in early-type galaxies has been acquired from some outside source, after the main stellar body was in place \citep{sarzi06,davis_misalign,bryant2019}.
Additional detailed work on the gas/dust ratios and associations with morphological disturbances suggests that most of the ``outside sources" are minor mergers with mass ratios $\gtrsim 10:1$  \citep{kaviraj2012,davis_minormergers}.
In some cases there are also additional clues about the origin of the gas from similarities or differences between the metallicities of the gas and the stars; for gas accreted in a minor merger, one might expect the metallicity of the accreted gas to be lower than the metallicity of the bulk of the stars \citep{griffith,davis+young}.

NGC~7465 clearly did acquire its cold gas from an external source, given the kinematic mismatches between the CO, \hi, and the stars.   But the gas probably did not come from a dwarf galaxy in a minor merger.  
The large \hi\ content \citep[9.5\e{9} \solmass;][]{serra12} is most consistent with the gas coming from a large spiral galaxy.  Indeed, there are several other similarly gas-rich galaxies still in its group \citep{li+seaquist,serra12}. 
Furthermore, the metallicity of the ionized gas in NGC~7465 is consistent with that of a spiral rather than a dwarf (Section \ref{cont}).
And finally, the inferred CO isotopic ratios in NGC~7465 are also consistent with it having captured its molecular gas from a spiral.  Its \thirco/\ceighto\ ratios are entirely unremarkable for spirals; its nuclear \tweco/\thirco\ line ratio is a factor of 2 to 3 higher than is common in spirals, but the measurement can be plausibly attributed to high temperatures and/or large linewidths rather than to abundance variations.

For other early-type galaxies with smaller gas masses, CO isotopic ratios may serve as an important complement to metallicity for identifying the original host of their external-source cold gas.  Or, indeed, for early-type galaxies with relaxed prograde gas, these indicators should help identify externally-sourced cold gas even when the kinematics are inconclusive.  As \thirc\ is a slow-release element with significant production in low- to intermediate-mass stars, it may retain signatures of an external origin longer than \twec\ or \eighto, and possibly also longer than kinematic disturbances (e.g.\ if the gas is now located in the interior of the galaxy where dynamical timescales are short). 

Figures \ref{fig:lir2} and \ref{fig:lir}, which are motivated by the discussions in \citet{taniguchi}, \citet{zhang2018}, and \citet{mendez-hernandez}, illustrate that 
ULIRGs and sub-mm galaxies tend to have high \tweco/\thirco\ line ratios $(\gtrsim 20)$ and low \thirco/\ceighto\ ratios $(\lesssim 2)$.  In fact, there also appears to be a strong correlation between these line ratios and the total FIR luminosity of a galaxy.  The cause of such a correlation is not obvious; presumably the total FIR luminosity of a galaxy is a proxy for other properties that are more fundamental.  \citet{zhang2018} argued it is a proxy for the IMF, though \citet{romano2019} cast doubt on that interpretation given the highly uncertain production rates of \eighto.

In any case, a thorough understanding of the chemical evolution of galaxies requires that we should be able to connect the sub-mm galaxies at $z \approx 2-3,$ at the highest luminosities in Figures \ref{fig:lir2} and \ref{fig:lir}, to their descendants in the local universe.  It is usually assumed that the sub-mm galaxies have actual abundance ratios [\thirco/\ceighto] $\approx$ 1 to 2, just like their observed line ratios, because their high \tweco/\thirco\ line ratios indicate that \thirco\ and \ceighto\ should be optically thin \citep[e.g.][]{danielson2013,sliwa_c18o}.  Over time the \thirc\ abundance should increase due to its production in low- to intermediate-mass stars.  And as the sub-mm galaxies already have stellar masses typically $10^{10}$ to $10^{11.5}$ \solmass\ at $z \approx 2-3$ \citep{boogaard2019}, their modern-day descendants will be amongst the most massive local galaxies.  But large galaxies in the local universe tend to have higher ratios of [\thirco/\ceighto] or [\thirc/\twec][\sixo/\eighto]  around 5 to 6 \citep[e.g.][]{martin2019,harada2018,romano2019}.
Whether the chemical evolution models can reproduce these kinds of variations on the appropriate timescales depends a great deal on their assumptions, particularly involving the production of the rare isotopes \citep{romano2019}.
Further isotopic observations of massive early-type galaxies, particularly those that might have retained their molecular gas through the cosmic star formation peak and through their transition to the red sequence, would be useful for testing the chemical evolution models.

On a related note, \citet{bayet12} and \citet{davis_cs} have argued that the [HCN/CO] abundance ratio might be a useful indicator of $\alpha$-element enhancement in molecular gas.  As this $\alpha$ enhancement is common in the stellar populations of early-type galaxies, studies of [HCN/CO] and [HCN/CS] abundance ratios (Section \ref{sec:cs}) could provide additional chemical clues to the evolution of early-type galaxies and the origin of their gas.

\section{Summary}

We present spatially-resolved molecular line observations of NGC~7465, an unusually gas-rich early-type galaxy that recently acquired $\approx 10^{10}$ \solmass\ of atomic and molecular gas in an interaction with another galaxy.  Its disturbed kinematics indicate that the gas is still in the process of settling and migrating inward.
We analyze ALMA observations of \tweco\ (1-0) at 0.\arcsec8 resolution (110 pc) plus the 3mm lines of  \thirco, \ceighto, \hcop, HCN, CN, \cch, and CS at $\approx$ 2\arcsec\ resolution (280 pc) and the continuum emission.
Besides NGC~7465, the data reveal an unidentified line source with a relatively bright peak line flux density of 2.5 mJy; it is probably a galaxy at $z = 0.2$ or 1.4.

We find two 3mm continuum sources in NGC~7465; the brighter one is a nuclear synchrotron source associated with the AGN that is also detected in low-frequency radio and hard X-ray emission.  The fainter one is associated with molecular clouds and star formation activity, traced by local peaks in \tweco, \thirco, HCN, \hcop,  and CS (though not CN).

The \tweco\ (1-0) distribution at 0.\arcsec8  resolution (110 pc) shows a low-inclination flocculent disk with a strong kinematic position angle twist spiraling inwards to a linear ridge or bar-like feature where the highest velocities are found.  In the inner 300 pc of the galaxy, the molecular gas kinematics are misaligned by $\approx$ 120\arcdeg\ with respect to the large-scale stellar rotation, by $\approx$ 100\arcdeg\ (in the other direction) with respect to the stellar kinematically-decoupled core, and by about 45\arcdeg\ with respect to the ionized gas kinematics.  
We conclude that the prominent stellar kinematically-decoupled core did not form out of the molecular gas present now; it must have had its origin in a previous event.
Furthermore, the complex misalignments may be signatures of outflows or other non-circular kinematics in ionized gas.
Despite the dramatic kinematic misalignments, there is no evidence of enhanced emission from the shock tracers SiO, \chhhoh, or HNCO.

All of the detected molecules are centrally concentrated except for \thirco\ (and possibly \ceighto\ and CS).
The distribution of \thirco\ has a prominent central dip of almost a factor of two in integrated line intensity and it peaks at about 300 pc from the nucleus.    
The central \thirco\ dip produces an unusually large nuclear \tweco/\thirco\ line ratio, $39 \pm 9$ at this resolution.  This ratio is higher than those found in typical spirals and early-type galaxies; it is comparable to those measured in ULIRGs and $z \approx 2-3$ sub-mm galaxies.
It also constrains the abundance ratio [\tweco/\thirco] $\geq 39 \pm 9,$ which is typical of spiral disks but is higher than sometimes found in spiral nuclei.  The isotopic ratios are thus consistent with the gas having been accreted from another spiral galaxy and transported rapidly (faster than the \thirc\ enrichment timescale) to the nucleus of NGC~7465.
The observed \thirco/\ceighto\ line ratio suggests [\thirco/\ceighto] = 4.0 $\pm$ 1.2, which is also consistent with a spiral galaxy origin but is significantly higher than corresponding ratios estimated in ULIRGs/sub-mm galaxies and lower than ratios in Local Group dwarfs.

Modest star formation activity is occurring in the center of NGC~7465 but there is no compelling reason to assume the intrinsic [\tweco/\thirco] abundance ratio is very different from the range of $40 - 60$ that is usually found in the disks of spirals.  In this case the \tweco\ (1-0) emission from the nucleus of NGC~7465 must be unusually optically thin, perhaps even having $\tau < 1.$  Such low optical depths are plausible because of (1) the high CO temperatures $\approx 100$~K inferred from single-dish data on multiple J-level transitions, and (2) the very large linewidth in the center of the galaxy.
The nuclear spectrum ($r < 140$ pc) has a FWHM of 250 \kms, a factor of 10 larger than elsewhere in the galaxy, due to the strong velocity gradient in the central misaligned molecular structure (possibly an edge-on circumnuclear ring).

HCN emission is relatively faint in NGC~7465 and \hcop\ is relatively bright, yielding HCN/\hcop\ ratios in the range of 0.4 to 0.6 everywhere we can measure -- these ratios are a factor of two lower than typical for spiral galaxies.  An assumption of LTE requires an intrinsic abundance ratio [HCN/\hcop] $< 0.25$ but even for densities too low to approach LTE the data still require [HCN/\hcop] $< 3,$ which is smaller than is found in some nearby spirals.
We also find (based on previously published nebular line fluxes) roughly solar metallicity in the ionized gas outside of the nucleus, where ionization is not dominated by the AGN.
Thus, even though the galaxy's gas-phase metallicity does not suggest an intrinsically low N abundance, the observed HCN/\hcop\ line ratios do still seem to require relatively low densities in the dense molecular phase (e.g.\ $n_{\rm H_2} \leq 10^{4.3}$\percc) and/or low HCN abundances.

We find no measurable gradient in the HCN/\hcop\ line ratio on scales of 100 pc to 1 kpc.  Thus, even though there is an AGN in NGC~7465, it is not affecting the J=1-0 line ratios of the dense molecular tracers on those scales (in marked contrast to some other AGN).  On the other hand, the proportion of dense relative to diffuse molecular gas is clearly changing with radius, as reflected by ratios like \tweco/HCN, \tweco/\hcop, and \tweco/CN.
Thus the physical properties of the densest phase of the molecular gas do not appear to change with radius, even though the properties of the more diffuse phase must be changing to reproduce the \tweco/\thirco\ variation.

The CN emission from NGC~7465 is optically thin and unremarkable in its intensity.  \cch\ is relatively bright, with \cch/HCN $\approx$ 1.  As \cch\ and \hcop\ have quite similar critical densities, we infer that the relative brightness of \cch\ is also an indicator of relatively low densities in the ``high-density" molecular phase.
The intensity and spatial distribution of CS emission in NGC~7465 are consistent with previous suggestions that CS can be enhanced in regions of star formation activity, here traced by low \oiiihb\ ratios.

All of the gas-phase data gathered about NGC~7465 to date -- the gas content, kinematics, metallicity, and molecular and isotopic abundance patterns -- are consistent with the interpretation that it acquired its gas recently from a large spiral galaxy rather than a dwarf galaxy.
In the broader context, however, 
more work is needed on isotopic abundance ratios in early-type galaxies.  Some of them are believed to have accreted their gas recently from a dwarf galaxy, while others may have retained small quantities of molecular gas from their previous lives as sub-mm galaxies at $z \approx 2-3,$ through the Universe's peak star formation epoch, all the way down to the present day.  More quantitative work is also required on modeling the isotopic abundance evolution of early-type galaxies, to test whether the current-day properties of these galaxies can be reproduced and what constraints they may impose on the evolutionary models.

\acknowledgements

We thank Davor Krajnovi\'c for helpful discussions on the stellar kinematics of barred galaxies.  This paper makes use of the following ALMA data: \\
ADS/JAO.ALMA\#2018.1.01253.S, \\
ADS/JAO.ALMA\#2016.1.01119.S, and  \\
ADS/JAO.ALMA\#2018.1.01599.S.
ALMA is a partnership of ESO (representing its member states), NSF (USA) and NINS (Japan), together with NRC (Canada), MOST and ASIAA (Taiwan), and KASI (Republic of Korea), in cooperation with the Republic of Chile. The Joint ALMA Observatory is operated by ESO, AUI/NRAO and NAOJ.
The National Radio Astronomy Observatory is a facility of the National Science Foundation operated under cooperative agreement by Associated Universities, Inc.

\vspace{5mm}
\facilities{ALMA}
\software{CASA \citep{casa}, Astropy \citep{astropy:2013, astropy:2018}}

\appendix

\section{A distant, unidentified source near NGC~7465}\label{app:ufo}

\begin{figure}
\includegraphics[scale=0.5,clip]{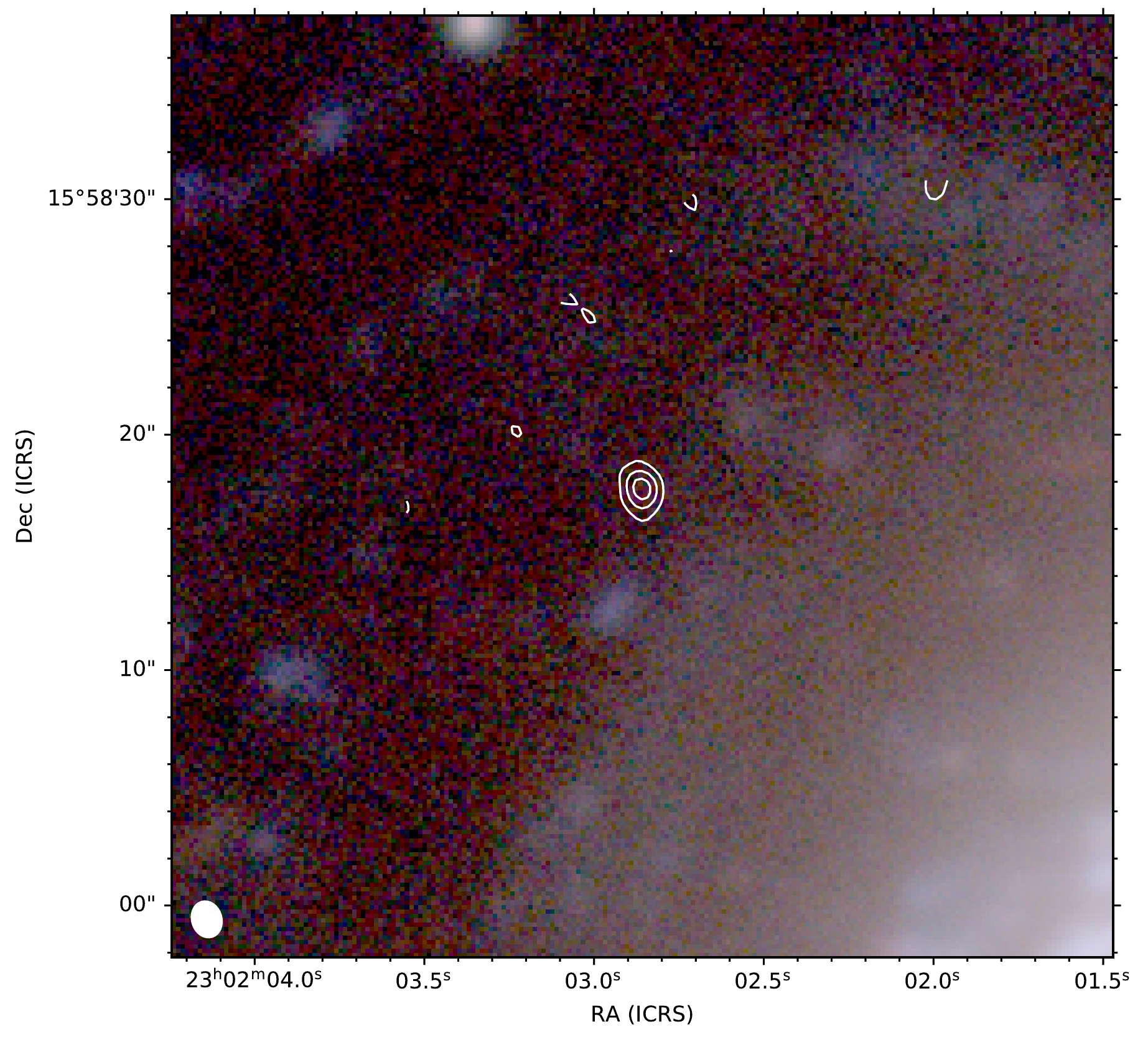}
\caption{Contours show the line emission from the unidentified source; levels are $(0.2, 0.5, 0.8)\times 3.12\times10^5$ (Jy bm$^{-1}$) Hz.  The background is the color composite of the MATLAS $u$, $g,$ and $i$ images \citep{duc15}. \label{fig:blob_overlay}}
\end{figure}

\begin{figure*}[]
\includegraphics[trim=0cm 7mm 0cm 5mm,clip]{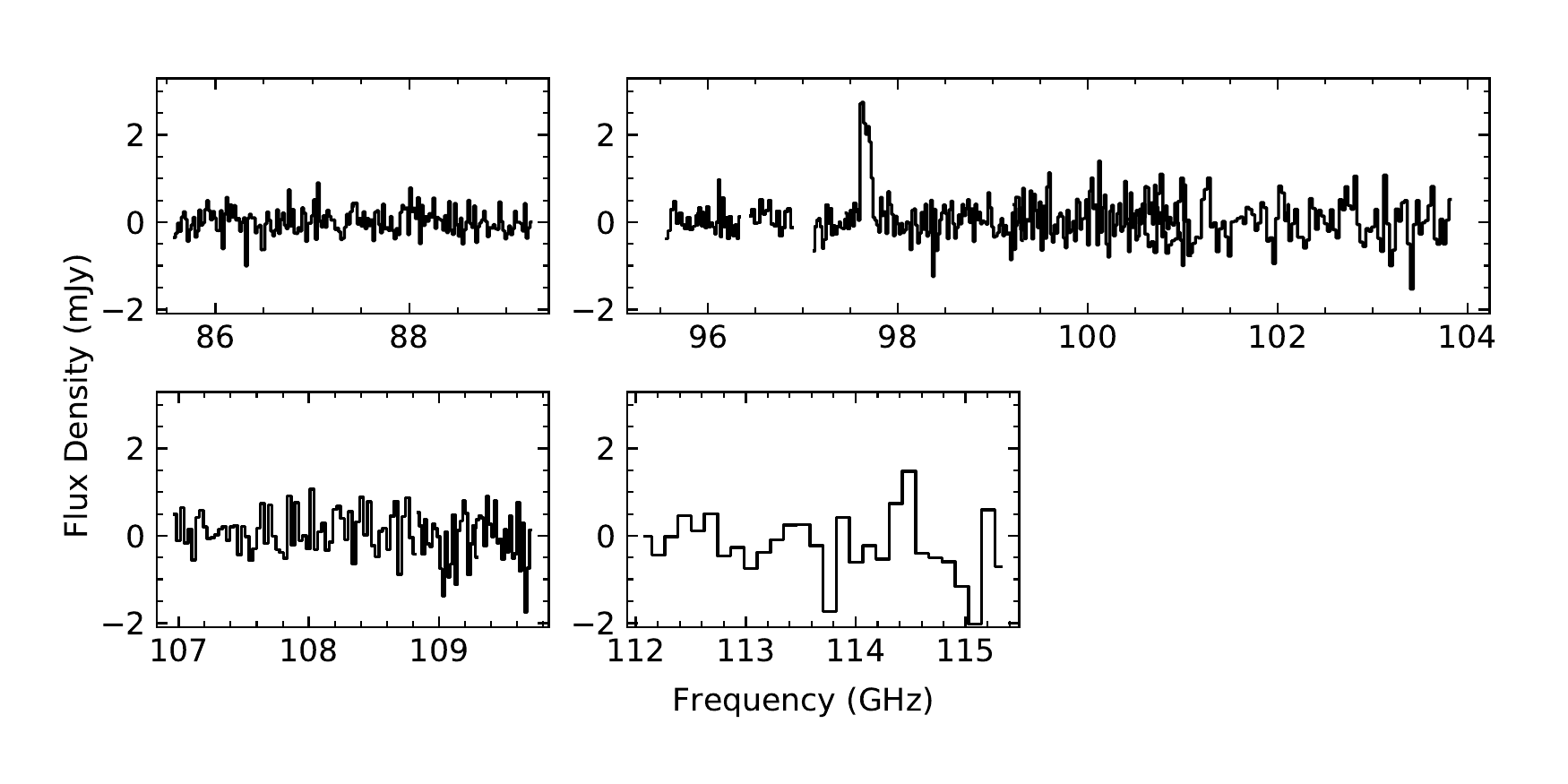}
\caption{3mm spectrum of the unidentified source.    Primary beam corrections have been applied.
\label{fig:blob_spec}}
\end{figure*}

We find strong line emission and 3mm continuum emission from an unidentified source near (in projection) to NGC~7465.
The line is centered at 97.67 GHz and its width is 0.14 GHz.
The source is modestly resolved in these data, and its position is 23$^h$ 02$^m$ 02.$^s$857, $+$15\arcdeg\ 58\arcmin\ 17.\arcsec8 (ICRS).  Figure \ref{fig:blob_overlay} shows its location on a deep optical image from the MATLAS survey \citep{duc15} and Figure \ref{fig:blob_spec} shows its spectrum. 
It also has a  3mm continuum flux density of (0.11 $\pm$ 0.01) mJy/beam and a slight suggestion of rotation along a northwest -- southeast axis, but the distance between the image centroids in the extreme channels is only 0.\arcsec35 so the rotation is not well resolved in these data.

Although the source is superposed on the outer parts of NGC~7465, where there are bright blue patches associated with recent star formation activity, there is no optical source coincident with the 3mm emission.  It certainly is not a part of NGC~7465 due to its large linewidth and the fact that its frequency does not match any known bright line.  But as we only have one detected spectral line in the frequencies covered by these data, it is difficult to identify the line.  If it is \tweco (1-0), it would be at $z=0.18.$  In this case \thirco(1-0) would be around 93.4 GHz, where we have no coverage, and HCN(1-0) would be beyond the low frequency end of ALMA Band 3.  If the line is \tweco(2-1), it would be at $z=1.3604$, \thirco(2-1) would be at 93.3 GHz, and HCN(3-2) would be at 112.6 GHz.  No line is apparent at 112.6 GHz but the data do not rule out typical HCN/\tweco\ ratios like those seen in local galaxies.  Similarly if the line is \tweco(3-2) at $z=2.5404,$ HCN(4-3) would be at 100.1 GHz but the limits on a nondetection there are not yet useful.
Firm identifications of the redshift in this case will probably require searches for the corresponding higher-J transitions of \tweco\ in the higher ALMA bands or the lower-J transitions at the Jansky Very Large Array.

\begin{deluxetable}{llccccccc}
\tablecaption{Flux, luminosity and mass estimates for the unidentified source\label{tab:Mdyn}}
\tablehead{
\colhead{Line ID} & \colhead{$z$} & \colhead{Diameter} & \colhead{Line width}  & \colhead{M$_\mathrm{dyn}$}  & \colhead{Flux} & \colhead{$L^\prime_\mathrm{line}$} & \colhead{M$_\mathrm{mol}$} & \colhead{M$_\mathrm{mol}$/M$_\mathrm{dyn}$} \\
\colhead{}             &    \colhead{}               & \colhead{(kpc)}       & \colhead{(\kms)}           & \colhead{(\solmass)}  & \colhead{(Jy \kms)} & \colhead{(K \kms\ pc$^2$)} & \colhead{(\solmass)} & \colhead{}
}
\startdata
\tweco(1-0) & 0.1802 & 1.06 & 409 & 5.2\e{9}/$\sin^2 i$ & 0.87 & 1.4\e{9} & 4.9\e{9} & 0.95 $\sin^2 i$ \\
\tweco(2-1) & 1.3604 & 2.94 & 217 & 4.0\e{9}/$\sin^2 i$ & 0.46 & 1.1\e{10} & 4.0\e{10} & 10.0 $\sin^2 i$ \\
\tweco(3-2) & 2.5404 & 2.81 & 115 & 1.1\e{9}/$\sin^2 i$ & 0.24 & 8.1\e{9} & 2.9\e{10} & 27 $\sin^2 i$ \\
\enddata
\tablecomments{Calculations are made assuming a flat universe with 
H$_0$ = 70 \kms~Mpc$^{-1}$, 
$\Omega_m = 0.3$,  and 
$\Omega_\Lambda = 0.7$.  The linear diameter is estimated from the emission centroids in the outermost channels.  The line luminosity $L^\prime_\mathrm{line}$ is calculated as in \citet{carilli+walter}.  Estimated luminosity conversions from J=2-1 or J=3-2 to J=1-0 are made using assumed excitations as in \citet{boogaard2019} and the molecular mass (with He) is then estimated from the inferred \tweco(1-0) luminosity using a conversion factor $\alpha = 3.6$ \solmass~(K \kms\ pc$^2$)$^{-1}$ \citep{boogaard2019}.}
\end{deluxetable}

Table \ref{tab:Mdyn} lists estimated line fluxes, luminosities, and masses for the most probable line identifications.  Assuming the object is in dynamical equilibrium, so that we can compare its inferred dynamical and molecular masses, it seems more likely that the line is \tweco(1-0) or \tweco(2-1) than any higher-J transition.  The higher-J levels would require the source to be quite close to face-on in order to make the inferred dynamical mass larger than the molecular mass.  The line profile is consistent with a rectangular or double-horned shape, suggesting it comes from a rotating disk with gas extending to the flat part of the rotation curve, so the dynamical equilibrium assumption is plausible by this measure.

The source is very similar in integrated line flux and width to the typical 3mm \tweco\ lines detected by ALMA in a blind survey of the Hubble Ultra-Deep Field \citep[HUDF;][]{gonzalezlopez,walter2016}.  Its continuum is brighter than is typical, as the brightest 3mm continuum source in the HUDF has a flux density of 46 $\pm$ 7 $\mu$Jy whereas this one has 110 $\pm$ 10 $\mu$Jy.
But overall, this source is compatible with being a similar object to those.  Most of them are identified as \tweco(2-1) at redshifts of 1.0 to 1.5, based on associations with optical counterparts and occasionally detections of higher-J lines in higher ALMA bands.  In fact, all of the \tweco\ detections in the HUDF have optical counterparts \citep{gonzalezlopez}, 
and they are typically 0.01 to 1 $\mu$Jy at an observed wavelength of 1 $\mu$m \citep{boogaard2019}.  The ground-based optical data shown in Figure \ref{fig:blob_overlay} cannot rule out the faintest of those typical optical counterparts.

\section{Possible evidence for an AGN-driven ionized outflow}\label{app:outflow}

\begin{figure*}
\includegraphics[width=\textwidth, trim=5mm 1.3cm 2cm 1cm, clip]{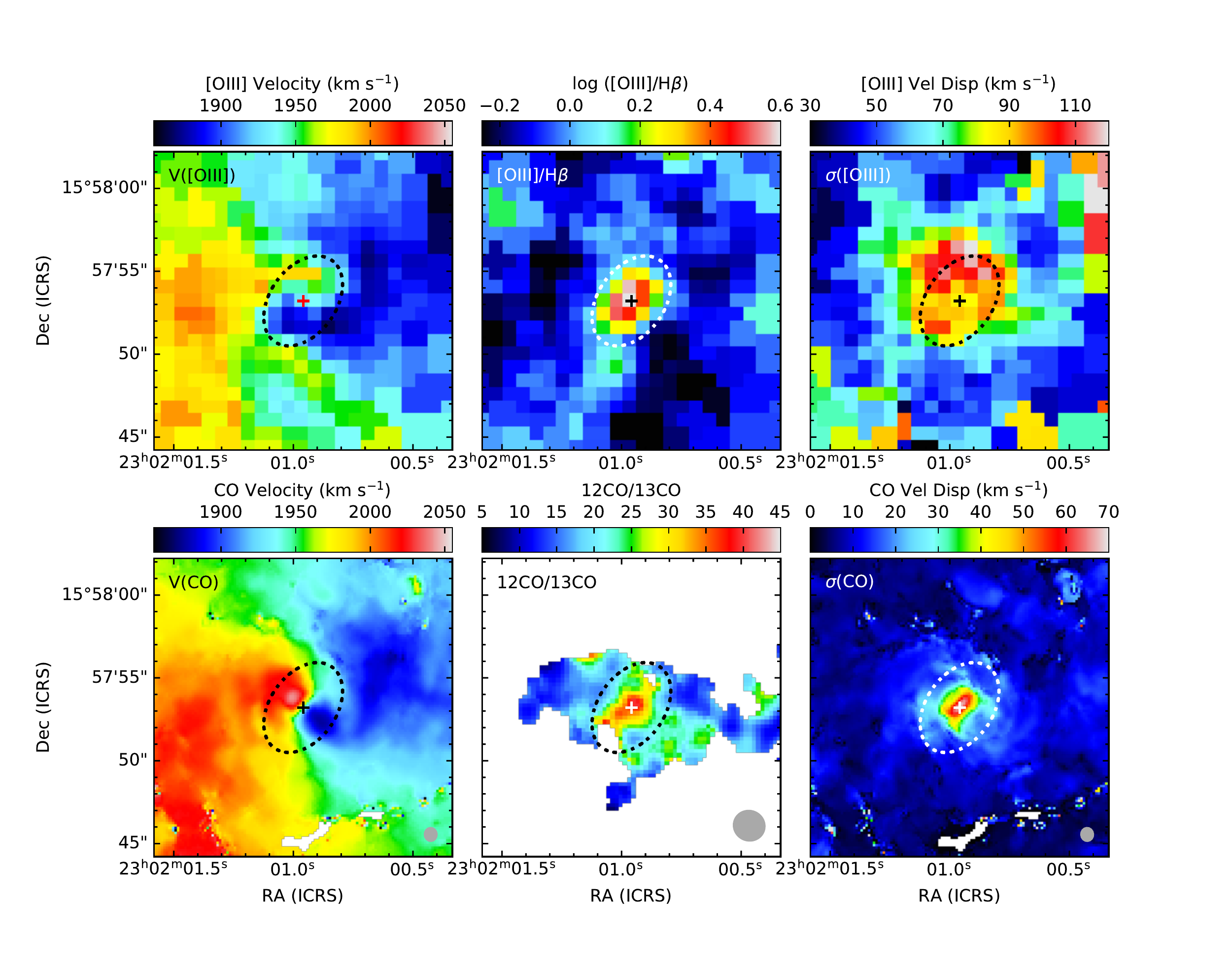}
\caption{The inner structure of NGC~7465.  The velocities and line ratios are the same data previously shown in Figures \ref{fig:7465cont}, \ref{fig:7465mom1}, and \ref{fig:7465_1213}; the panels on the right side also show the velocity dispersions in \tweco\ and \oiii.  Beam sizes for the ALMA data are shown as ellipses in the lower right corners.  The cross marks the location of the nuclear 3mm continuum source, and the dashed ellipse (6\arcsec\ $\times$ 4\arcsec, 850 $\times$ 570 pc) roughly indicates the region where possible outflow signatures can be seen in the optical data.
\label{fig:outflow}}
\end{figure*}

Figure \ref{fig:outflow} shows a side-by-side comparison of the inner structure of NGC~7465, in some indicators that might reveal a small AGN-driven ionized outflow.  \citet{ferruit} presented narrowband HST \oiii\ and H$\alpha$ imaging which showed enhanced \oiii/H$\alpha$ in a $r \lesssim 2$\arcsec\ region with a southeast-northwest elongation, consistent with the structure in the \atlas\ \oiiihb\ data.   This region of more strongly-ionized gas is expected to trace enhanced shocks and/or harder radiation fields.  The velocity of the ionized gas shows a gradient along a similar kinematic axis of roughly $-$35\arcdeg, as indicated by the ellipse in Figure \ref{fig:outflow},  and the velocity dispersion in \oiii\ is also enhanced near the ends of the major axis of this ellipse.  However, the velocity resolution of the \oiii\ kinematic data is not good enough to show any signatures of double-peaked line profiles.

The \tweco\ kinematic and photometric position angles in this region show that the molecular gas predominantly traces a disk or a ring perpendicular to the suggested outflow axis.  Thus, enhanced \oiiihb\ ratios along that axis might simply reflect lower opacities on that axis and higher opacities in the dusty molecular disk, rather than actual outflow.  The enhanced \tweco\ velocity dispersion along the proposed outflow axis is probably then related to the kinematics of the molecular disk rather than the outflow, and enhanced \tweco/\thirco\ line ratios along the proposed outflow axis might reflect higher temperatures or larger linewidths (Section \ref{sec:disc_isotop}).  In short, the
features observed here could be caused by an ionized outflow but are not definitive evidence; higher resolution spectrosocopy targeting the optical nebular lines would be useful in testing for the existence of an outflow.

\section{Line fluxes in NGC~7465.}\label{app:bigtab}

Table \ref{tab:bigtab} presents all the line fluxes measured in the regions used for the analysis of spatial variations in the molecular properties of NGC~7465. 

\movetabledown=5.5cm
\begin{rotatetable}
\begin{deluxetable*}{lrrrrrrrrrrrrr}
\tablecaption{Line fluxes in the defined regions of NGC~7465\label{tab:bigtab}}
\tablewidth{0pt}
\tabletypesize{\scriptsize}
\tablehead{
\colhead{Rgn} & \colhead{Dist.} & \colhead{Area} & \colhead{\tweco} & \colhead{\thirco} & \colhead{\ceighto} & \colhead{HCN} & \colhead{\hcop} & \colhead{CN} & \colhead{CS} & \colhead{\cch} & \colhead{\chhhoh} & \colhead{SiO} & \colhead{HNCO} \\
\colhead{ID} & \colhead{(\arcsec)} & \colhead{($\square$\arcsec)} & \colhead{(\jykms)} & \colhead{(\jykms)} &
\colhead{(\jykms)} & \colhead{(\jykms)} & \colhead{(\jykms)} & \colhead{(\jykms)} & \colhead{(\jykms)} & \colhead{(\jykms)} & \colhead{(\jykms)} & \colhead{(\jykms)} & \colhead{(\jykms)}
 \\
}
\startdata
\input{rgn_linefluxes.tex}
\enddata
\tablecomments{Regions are identified by number in Figure \ref{fig:regions}.  The distance in column 2 refers to the typical distance between the nuclear continuum source and points within the region, and it indicates the radial coordinate used for each region in Figures \ref{7465radial1} and \ref{7465radial2}.  Region 9 is a much larger irregular area roughly 38\arcsec\ by 32\arcsec\ in diameter, encompassing virtually all of the detected \tweco\ emission in these data, and its typical radial coordinate is arbitrarily set to 10\arcsec\ for ease in plotting. Line ratios measured in K units can be calculated from these table entries by multiplying the ratio in Jy units by the square of the ratio of the lines' wavelengths.}
\end{deluxetable*}
\end{rotatetable}

\bibliography{alma7465}



\end{document}

%% file: rgn_linefluxes.tex
1 & 0.5 & 3.1 & 3.515 (0.040) & 0.096 (0.013) & $<$ 0.039 & 0.132 (0.012) & 0.303 (0.013) & 0.226 (0.006) & 0.041 (0.009) & 0.091 (0.014) & $<$ 0.035 & $<$ 0.040 & $<$ 0.037 \\ 
2 & 2.3 & 8.8 & 6.877 (0.069) & 0.325 (0.023) & $<$ 0.061 & 0.128 (0.022) & 0.219 (0.018) & 0.122 (0.015) & 0.087 (0.018) & 0.104 (0.022) & $<$ 0.057 & $<$ 0.066 & $<$ 0.061 \\ 
3 & 1.9 & 5.9 & 4.852 (0.062) & 0.186 (0.020) & $<$ 0.050 & 0.092 (0.019) & 0.222 (0.018) & 0.162 (0.008) & $<$ 0.041 & $<$ 0.065 & $<$ 0.057 & $<$ 0.053 & $<$ 0.058 \\ 
4 & 3.4 & 7.4 & 3.787 (0.053) & 0.182 (0.018) & $<$ 0.054 & 0.045 (0.014) & 0.109 (0.015) & 0.058 (0.007) & $<$ 0.043 & 0.048 (0.016) & $<$ 0.042 & $<$ 0.047 & $<$ 0.050 \\ 
5 & 2.9 & 57.2 & 30.888 (0.277) & 1.266 (0.093) & 0.311 (0.091) & 0.477 (0.070) & 1.122 (0.070) & 0.814 (0.035) & 0.337 (0.068) & 0.468 (0.086) & $<$ 0.260 & $<$ 0.234 & $<$ 0.222 \\ 
6 & 5.8 & 10.0 & 2.163 (0.047) & 0.130 (0.014) & $<$ 0.041 & $<$ 0.041 & $<$ 0.034 & $<$ 0.018 & $<$ 0.034 & $<$ 0.036 & $<$ 0.027 & $<$ 0.039 & $<$ 0.032 \\ 
7 & 7.9 & 15.6 & 4.312 (0.091) & 0.201 (0.029) & $<$ 0.082 & $<$ 0.057 & $<$ 0.067 & $<$ 0.037 & $<$ 0.053 & $<$ 0.069 & $<$ 0.060 & $<$ 0.064 & $<$ 0.073 \\ 
8 & 5.3 & 3.1 & 0.598 (0.016) & 0.039 (0.005) & $<$ 0.013 & 0.019 (0.005) & 0.040 (0.005) & $<$ 0.007 & $<$ 0.013 & $<$ 0.014 & $<$ 0.015 & $<$ 0.017 & $<$ 0.018 \\ 
9 & 10.0 & 759.8 & 80.792 (0.866) & 2.540 (0.279) & 0.845 (0.265) & $<$ 0.668 & 1.328 (0.212) & 0.843 (0.131) & $<$ 0.643 & $<$ 0.602 & $<$ 0.659 & $<$ 0.631 & $<$ 0.745 \\ 